# Understanding the Prevalence of Caste: A Critical Discourse Analysis of Caste-based Marginalization on X

NAYANA KIRASUR, Rutgers University, US
SHAGUN JHAVER, Rutgers University, US

Despite decades of anti-caste efforts, sociocultural practices that marginalize lower-caste groups in India remain prevalent and have even proliferated with the use of social media. This paper examines how groups engaged in caste-based discrimination leverage platform affordances of the social media site X (formerly Twitter) to circulate and reinforce caste ideologies. Using a critical discourse analysis (CDA) approach, we examine the rhetorical and organizing strategies of 50 X profiles representing upper-caste collectives. We find that these profiles leverage platform affordances such as information control, bandwidth, visibility, searchability, and shareability to construct two main arguments: (1) that their upper caste culture deserves a superior status and (2) that they are the "true" victims of oppression in society. These profiles' digitally mediated discursive strategies contribute to the marginalization of lower castes by normalizing caste cultures, strengthening caste networks, reinforcing caste discrimination, and diminishing anti-caste measures. Our analysis builds upon previous HCI conceptualizations of online harms and safety to inform how to address caste-based marginalization. We offer theoretical and methodological suggestions for critical HCI research focused on studying the mechanisms of power along other social categories such as race and gender.

## 1 Introduction

Computing technologies such as the internet have been viewed as an equalizing infrastructure with the potential to interrupt inequalities, including caste, a social hierarchy in India and its diasporas. The rise in computing power and internet connectivity is believed to bring economic development that can erode the day-to-day authority of caste [103]. Techno-optimists also assert that anyone who can operate mobiles and access the internet can now participate as equals on social media, a space usually envisaged as a public sphere [106, 123]. However, though mobile and internet penetration in India has increased [69], and social media has been used in innovative ways to challenge the caste hierarchy [65], the hold of caste has not weakened. Lower-caste communities continue to be relegated to the margins politically, socially, economically, and culturally [14, 27, 37, 119, 125, 132]. Upper-caste norms, traditions, and values are repeatedly normalized and even held superior to lower-caste cultures by social institutions such as mainstream media, legal systems, educational spaces, and family.

On social media platforms, caste-based hate speech and calls for violence against lower castes remain prevalent [15, 21, 82, 112, 126]. Caste-positive[1] individuals and groups use these platforms to express caste superiority and build caste-based affiliations. Such online activities inherently involve structural power imbalances but often evade critical scrutiny. In this paper, *we focus on*

[1]Shanmugavelan [122] writes that dominant castes regard and portray caste hierarchy as a positive culture, which he terms as caste-positivism. Throughout this paper, we use the term *caste-positive* to refer to the practice of favoring the caste hierarchy.

Authors' Contact Information: Nayana Kirasur, nayana.kirasur@rutgers.edu, Rutgers University, US; Shagun Jhaver, shagun.jhaver@rutgers.edu, Rutgers University, US.

*the ideological underpinnings and organizing mechanics of dominant-caste community accounts on the social media site X (formerly Twitter) and underscore the role of social media in reproducing caste hierarchy.* By 'community accounts,' [2] we mean user profiles that are not associated with an individual but represent a collective, an issue concerning a social group, or an organization.

The prevalence of caste has been a topic of inquiry in several fields, including sociology, cultural studies, and political science. Some of the questions scholars have asked are: How does caste manifest in contemporary India? How does caste get re-entrenched in our social milieu? [10, 60, 128, 138]. Given that much of our everyday life is now digitally mediated [102], the fields of CSCW and social computing are uniquely suited to explore the reproduction of caste hierarchy online. Recent research in these areas, for instance, examines the role of caste in shaping Indian politicians' social media use [136, 137], exclusions of women belonging to marginalized castes from India's #MeToo movement [101], and the impact of caste relations on gig work [4]. We contextualize our work within this prior literature to investigate online articulations of upper caste ideologies. We also extend conversations already underway in social computing and allied fields that explore mechanics of power along the lines of gender, race, sexuality, etc., and examine patterns of marginalization in social media use across varied socio-geographical contexts [131].

To identify how caste practices manifest and are sustained on social media, our study examines discourse on X, a platform where significant political and cultural work occurs [11–13, 58, 91, 111]. We frame our research around X profiles of caste-positive communities for two reasons: First, drawing upon boyd and Heer [24], we see profiles as ongoing conversations that are rich information sources about how communities leverage the diverse platform affordances, including searchability, visibility, shareability, and information control, to express their caste ideologies and organize around their caste group's interests. Second, caste ideologies and caste-based interests tend to be expressed in a focused manner on community profiles as opposed to individual accounts, where caste expressions may be intermittent and scattered. Community profiles on social media are sites where we can observe ideologies in the making as well as ideologies in action.

As such, our paper builds on recent research that investigates online networking and cultural practices among dominant or structurally privileged groups, such as white supremacists [38, 63], men's rights activists [44, 45, 62, 107], and other far-right groups in various countries such as the United Kingdom [6], Germany [114, 118], the United States [96, 127, 129], and India [53]. We present evidence on how online *discursive strategies* of caste-positive community accounts on X diffuse caste ideologies and reconstitute caste identities. Here, we draw upon Wodak's [142] understanding of discursive practices, who defines them as ways in which language and content (speech or media) are employed in social environments to serve specific purposes. Some of the examples of discursive practices she outlines are (1) argumentation or the use of reasoning, rhetorics, claims, and counterclaims to justify or legitimize particular views or ideologies, (2) referential strategy or the construction of in-groups and out-groups, and (3) predication or the construction of certain groups positively or negatively. These strategies, as we will see in the findings section, overlap and work in tandem to help produce and reproduce unequal power relations [64, 139, 143] between lower and upper castes. The research questions that guide this paper are:

**RQ 1:** How do caste-positive communities leverage platform affordances to voice their caste interests and organize around them?

**RQ 2:** What discursive strategies do caste-positive community accounts on X use to communicate and promote their caste ideologies?

[2]Not to be confused with X communities, a feature that creates a dedicated space for only the member accounts to participate in discussions.

Using a Critical Discourse Analysis (CDA) approach, we study 50 caste-positive community profiles on X. Our analysis shows that in this unique socio-technical environment, platform affordances (especially information control, searchability, visibility, bandwidth, and shareability) enable accounts to simultaneously deploy both rhetorical strategies and organizing strategies in ways that produce two complementary argumentative narratives. First, the accounts use various tactics to assert that their caste cultures are superior. As the paper explains further in Section 4.1, they do so by building spaces dedicated to upper caste narratives on X, constructing a caste-coded and Hindu audio-visual culture, undermining anti-caste struggles, and translating caste networks into caste capital. The cumulative effect of caste-positive spaces, the assertions of caste pride, and the consolidation of caste-based affiliations contribute to the strengthening and circulation of upper-caste cultures and discourses.

Second, the accounts construct another narrative that claims that the dominant castes are the "true" victims. As we detail in 4.2, tactics to advance this narrative include appropriating resistance mechanisms of the historically oppressed, building evidence of victimhood, denigrating progressive policies that support lower castes, and expressing nostalgia for the so-called "lost" cultures and traditions. Through these tactics, the rhetoric of anger and resentment toward constitutionalism and the lower castes combines with the rhetoric of revivalism. This, in turn, enables these accounts to mobilize their caste communities to strike down egalitarian efforts and continue practices that keep caste and even gender hierarchies intact.

This paper underscores the increasing critical and humanistic emphasis within Human-Computer Interaction (HCI) that investigates iterations of privilege and power [18]. Our research makes three main contributions: (1) We expand the current scholarly conversation around how dominant-caste groups consolidate caste power on social media. Our analysis contributes to the emerging approach of Critical Caste and Technology Studies (CCTS), which offers a "communication-, media-, and technology-based critique" of contemporary manifestations of caste [121]. (2) We demonstrate how the strategic use of critical and qualitative approaches to study profiles of online communities can provide a nuanced understanding of how structurally privileged groups reproduce hegemonic cultures via social media. (3) We highlight the wide spectrum of discriminatory content and push for an expansive understanding of online harm and safety.

## 2 Related Work

### 2.1 Workings Of Caste

Caste is a system of social hierarchy that ascribes one's status at birth and dictates much of life in India and among its diasporas. It is characterized by graded inequalities [2] that produce social, political, and economic dispossession of those who occupy the bottom rungs of the structure. A Brahmanical[3] understanding of caste is linked to the notions of *varna,* which is a four-fold hierarchical classification codified in the Hindu religious scriptures, and *jaati*, a term that describes a range of closed groups that tend to be organized locally. Castes that fall within the four-fold varna hierarchy are the Brahmins, Kshatriyas, Vaishyas, and Shudras, who are also referred to as Savarnas — literally meaning 'those with varna,' i.e., they fit within the varna classification system. The most marginalized are India's Tribal and Dalit (translates to 'to be broken' or oppressed) populations; these communities are the Avarnas, which are excluded from the varna system. This system heavily regulates individuals' social interactions since inter-caste contact, dining, and marriage are discouraged.

---

[3]Brahmanical or Brahminism (noun.) refers to the sociopolitical ideology that vests Brahmins with a superior status, power, and privilege.

While unpacking the caste hierarchy, Omvedt [104] refers to varna as an ideological classification that doubles as a hierarchy of purity and pollution. The Brahmins hold the most social and cultural power as priests, administrators, and very often landlords and bureaucrats; the Kshatriyas are considered to embody strength since they were historically the rulers and warriors; the Vaishyas are the trading class; and the Shudras have been relegated to service roles of peasants, artisans, and toilers. Outside the hierarchy are the Dalits, who "[perform] tasks both laborious and polluting (carrying away dead animals, keeping the village clean, the most menial agricultural labor, and service to village dominant castes and bureaucrats); Dalits most frequently [live] in separate settlements outside the village itself" (p. 8). Since they tend to be restricted to the "polluting" jobs, they were historically considered the untouchables.

While the varna system plays a role in determining one's status, such a monolithic view of caste tells us only half the story. A critical perspective on caste recognizes that this structure is not a monolith but deeply contextual in nature. This means that on-the-ground realities of how caste hierarchy plays out are largely shaped by historical, socio-political, cultural, and economic conditions with complex inter- and intra-regional variations [124]. In response to the highly contextual and historical nature of caste oppression, innumerable anti-caste struggles have taken place throughout history and continue to emerge locally [104]. These struggles are pluralistic and are demonstrated not just as overt forms of "contention" [133] such as protests and strikes but also embedded in communities' everyday routines and spaces, including online [130], that challenge caste-based inequalities.

The Indian constitutional and legal-penal framework, effectuated in 1950, prevents caste-based discrimination and identifies lower castes as a protected category. This enforcement arose after a long anti-caste struggle under the leadership of Jotirao Phule and Savitribai Phule in the 1800s and Periyar (who led the Dravidian movement in the south) and Bhimrao Ambedkar in the early to mid-1900s. After 1947, when the newly independent country of India failed to keep the promise of eliminating caste or even ameliorating caste oppression, another movement rose with the emergence of Dalit Panthers in the 1970s, who fought against the Savarnas. The subsequent anti-caste movement attempted to consolidate middle non-Brahmin castes against Brahminical hegemony (Brahminism), as seen, for instance, in the efforts of anti-caste thinkers such as Kanshi Ram and Sharad Patil, who insisted on a caste-class analysis of society and organized an alliance between Dalits and Shudras against Brahminism. This led to the expansion of the affirmative action policy to include not just Dalit and tribal communities but also other lower castes under the umbrella of "Other Backward Classes" or OBCs [104] (pp.73-75). The current anti-caste movements emphasize gaining political power and driving social action led by the marginalized.

As we will see in the rest of the paper, these historical and contextual specificities of caste and anti-caste movements greatly shape contemporary realities of caste, including on social media platforms such as X.

### 2.2 Caste as a Function of Culture, Space, and Discourse

Despite constitutional measures and widespread social counter currents, caste-based exclusion continues. It dictates one's social status, whom one can marry, what occupations one can or cannot do, and access to one's resources and networks. To understand how these forms of discrimination and exclusion are operationalized, we draw upon the idea of structural power and marginalization. We use D'Ignazio and Klein [39]'s framing of power that argues that our social institutions and systems enable some groups unearned advantages, whereas other groups experience systematic disadvantage or marginalization because those very institutions are not meant to work for the latter groups. In this case, the structure of caste privileges upper-caste communities and marginalizes lower castes.

While tying this idea of structural power to caste, Kumar [92] contends that contemporary manifestations of caste can be understood as upper caste groups' ability to cumulatively accrue social, economic, and cultural capital while excluding the lower castes. Drawing from this, we frame social categories as systems of social stratification characterized by groups possessing structural power that enables their domination.

Integral to caste are shared *cultures*, which attempt to solidify upper-caste members' superior place in the social order. Such a mechanism of maintaining caste is associated even with its genesis. Ambedkar [2] writes that caste can be understood as social units that raised an enclosure around themselves and closed themselves off from other groups. These units not only encouraged deep cultural unity among themselves but also discouraged social relations with members of other units. While surveying the relationship between experience, space, and justice, Guru [59] argues that caste hegemony plays out through *spaces* as well. Since ideologies cannot be articulated on their own, exclusive spaces act as a vital backdrop for expressions of caste. Similarly, caste is also maintained through homophily and shared *discourses*. Bairy [10] refers to caste as an "intersubjectively made reality" that is articulated through "symbolic, non-verbal, and verbal communicative modes" [122] that constitute almost all social interactions, including those that occur online.

Pathania and Tierney [108]'s ethnographic analysis of caste dynamics at an Indian university confirms this by showing that upper-caste students tend to pursue homophily and not befriend lower-caste students (p. 10). Even today, upper-caste communities prefer partners from within their caste or from a similar "social background," a euphemism for caste. This is evident through the caste-specific preferences expressed on matrimonial websites [100, 109]. The strong association between caste, occupation, and mobility is another peculiar feature of caste since lower castes historically inherited low-paying and service-oriented occupations and the lack of attendant privileges. Even when they manage to attain occupational mobility — by virtue of their efforts and affirmative action policies — social mobility within the hierarchy is not guaranteed [105]. This significantly impacts the experiences of lower castes and how they navigate their workplaces. Vaghela et al. [135] discuss how Dalit engineers navigate the computing industry by tactically leveraging social media to find other Dalit engineers. At the same time, these engineers rely on anonymity or curate their online network carefully to keep their caste identity hidden and avoid discrimination in their workplace. With this paper, we build on these prior efforts to understand how caste intersects with computing cultures.

### 2.3 Group Interactions and Platform Affordances

Given that the practice of caste generates social and cultural capital, it inherently involves the process of displaying, identifying, marking, and engaging with caste-based identities. Using Goffman [57]'s theory of presentation of self, Beckmann and Gross [16] analyze how social media provides a space for participants to converse akin to a stage performance. With respect to collaborations in particular, they say that members of a group typically present individual performances that contribute to a shared goal. In this research, we draw from these analyses to examine how caste-positive community accounts on X contribute to reinscribing caste.

In a sociotechnical environment such as social media, identity-based expressions and collaborations are influenced by platform affordances. The concept of affordance by Gibson [54] has been welcomed and adapted extensively within the HCI research. Kaptelinin and Nardi [84] explain that the crux of the affordance theory lies in the understanding that "it is our basic nature to see the environment in terms of the range of action possibilities (i.e., affordances) offered to us by the objects in the environment." They retain this idea and reconceptualize it for HCI by defining the environment in cultural and technological terms. Calling it the "mediated action perspective on affordances," they write that our "activities and minds are mediated by culturally developed tools, including technology." While their paper focuses on individuals interacting with computing technologies, our paper considers caste-positive collectives' usage of social media.

Another useful insight from Gibson is that affordance refers to both the socio-technological environment *and* the actor. Evans et al. [42] propose a set of criteria that can be employed to define technological affordances and analyze them in relation to platform features and the various outcomes of technology use. Their conceptual framework lies at the intersection of the technology object or feature, the actor, and their specific contexts, making it a helpful lens through which to explicate the "relational and situated nature" of technology use. For a detailed glossary and explanation of affordances, see Evans et al. [42] as well as Fox and McEwan [48].

On X, features such as profile pictures, header photos, and bios reveal (or hide) information about profiles' preferred identities and attitudes. These features enable affordances [48] such as information control (choices that determine what others see or don't see) and visibility (the degree to which posts are visible to others, whether intended or not). In addition, reposts, replies, followers, and following lists indicate connections or invitations for connections between accounts, facilitating the outcome of homophily seeking. Other features on X enable searchability (one can find other people by searching names and find content by searching keywords), persistence (one can re-up older posts to initiate dialogues), and synchronous or asynchronous conversations; these are affordances that determine how identities (in this case, caste-based identities) interact with each other [83]. We study how affordances such as information control, visibility, shareability (the ability to share content [26] with users), searchability, and bandwidth (affordance indicating the breadth of social cues available on the platform) intersect with caste-positive expressions on X.

### 2.4 Communities of Power on Social Media and Online Harm

Recent studies show how dominant identities leverage social media's affordances and "participatory culture" [71] to organize around their interests, seek buy-ins from their in-group members, and share malicious content against the marginalized. Previous scholarship in HCI and social computing has examined how certain groups, such as women [113], LGBTQ+ individuals [115], and racial minorities [66, 144], are more vulnerable to online content-based harm.

In the Indian context, Mankekar and Carlan [97] show how calls for violence against those who critique the Bharatiya Janata Party (or BJP, a Hindu nationalist party) government are driven by nationalism and constructed by emotionally charged articulations of Hindu nationalists online. Bhimdiwala et al. [20] discuss how deep-seated Islamophobia and patriarchy in India contributed to the networked harassment case of Sulli Deals and Bulli Bai, wherein Indian Muslim women were auctioned on applications hosted on GitHub. Conducting a discourse analysis of social media posts about Hindu nationalism, Bhatia [19] found that online discussions about nationalism in India often constitute hate speech to reinforce the Hindutva ideology that marginalizes minoritized groups, especially Muslims. This strand of literature makes it clear that structurally privileged communities' discursive presence and organizing efforts online contribute to both individual and collective harm.

While developing a framework to analyze online harm, Scheuerman et al. [116] write that online harm can be of four types — physical, emotional, relational, and financial. Applying this framework to the content circulated by dominant groups, we can infer that a constant messaging that favors dominant groups combined with malicious content, hate speech, and disinformation against the marginalized amounts to significant emotional and relational harm at the least. The scale of the harm is high since the target is an entire community that is already vulnerable and has restricted agency in taking mitigating action. It is also concerning because the intent of the harm is usually to (re)assert dominance and suppress resistance on a public platform.

To contribute to the literature that characterizes structurally powerful groups' online behavior, we examine community accounts that explicitly represent the interests of a dominant caste community/group(s), particularly the Brahmins, Kshatriyas, and dominant middle castes, and investigate their patterns of use on X. Through this research, we intend to expand the conversation about dominant communities and their mechanics of power by focusing on the category of caste. We hope that this study will trigger additional conversations on how we can frame marginalization of oppressed groups as a manifestation of online harm that all stakeholders, including platforms, the state, and civil society, must address.

## 3 Methods

Gathering social media data to study discriminatory content has been challenging for the scholarly community [32, 34, 98]. On one hand, platforms' control over application programming interfaces (APIs) hinders data collection processes, restricts researchers from sharing data, and discourages the replication of studies [35, 50]. For example, Elon Musk, the CEO of X, recently put the platform's API behind a paywall [28], making large-scale data from the website inaccessible for research. On the other hand, discriminatory content tends to be concealed, making it difficult for researchers to find. Marks and Stanfill [98] discuss such challenges vis-à-vis racist speech and argue that methods such as hashtag-tracking and keyword searching are insufficient sampling techniques since discriminatory content does not always include overt hate vocabulary.

Given this context, to collect our data, we first developed inclusion criteria that informed how we sampled the profiles. Since our focus is on caste-positive community accounts on X, we decided to gather accounts that explicitly expressed caste-positive attitudes through their profile characteristics, posts, and reposts. We began by familiarizing ourselves with caste-focused conversations on X, reviewed literature [82, 122, 126] using similar approaches to identify relevant data, and engaged in discussions among ourselves to arrive at the following inclusion criteria:

1. Accounts that represent and cater to a particular dominant caste community (*jaati*) or caste group (*varna*).
2. Accounts that voice the interests of dominant castes.
3. Accounts with 4,000 or more followers.

While we strictly adhered to criteria 1 and 2, we made exceptions for criterion 3 to include 10 profiles that we had already identified in our preliminary observation as sufficiently theoretically significant to answer our research questions despite having slightly fewer than 4,000 followers.

### 3.1 Data Collection

Given the aforementioned constraints of sampling social media data for qualitative analysis, we executed our data collection process in two phases, discussed below. We primarily drew upon two of Marks and Stanfill's [98] profile-based sampling methods: (1) recursive searching (mining keyword and hashtag search results for additional keywords) and (2) algorithmic recommendation (following the platform's algorithmically suggested content).

*Phase 1*

To gather data, we first made a new X account (*study account*, henceforth). Based on prior literature [122, 134] and our familiarity with caste discourse on X, we enlisted a preliminary set of keywords and hashtags that are frequently used by dominant castes to express caste ideologies. These included "Brahmin," "Kshatriyas," "#ProudBrahmins," "#Rajputboy," and '#unreserved." We collected posts containing these keywords to constitute our initial dataset. Next, we reviewed the accounts that published these posts and selected those that met our selection criteria. After selecting each profile, we gathered profile-specific data, which included what Crosset et al. [34] call an "assemblage of digital traces:"

1. Profile information (bio, profile picture, header photo)
2. Latest 30 posts
3. Online interactions, such as replies under posts and quote posts
4. Audio-visual content in the collected posts

We limited data collection to the profile information and the latest 30 posts since our preliminary analysis (review of profiles coupled with iterative discussions between the authors) of the first 15 profiles showed that examining 30 posts of each profile was sufficient to understand the caste community the profiles were catering to, the caste-based interests (often mentioned in the profile information) they were promoting, and their goals and key strategies (mentioned in the bio or observable in their latest posts). The rationale behind gathering only the latest posts was to focus on the accounts' current caste articulations.

*Phase 2*

In Phase 2, we expanded our sample using three strategies:

1. For each profile, we examined the first 50 followers and following accounts as shown (i.e., determined by X's account ranking algorithms) on that profile's followers and following pages. We included those that met our criteria.
2. We examined the profiles whose content was algorithmically shown to our study account in its regular news feeds and selected those that satisfied our criteria.
3. As we started memoing the profiles, we found additional keywords and hashtags that we fed into our search strategy to find additional profiles.

Repeating phases 1 and 2 and leveraging X's feed algorithms for the data collection process helped us trace caste-positive communities. The search queries were conducted in languages that the authors were familiar with: English, Hindi, Marathi, and Kannada. We sampled a total of 50 profiles between September 2023 and March 2024, after which we determined that the data collected was sufficient to characterize upper-caste communities' discursive strategies comprehensively. We first listed the 50 profiles on a spreadsheet and assigned a number to each — P1, P2, ..., P50. Next, we used Meltwater, a global media monitoring and analysis platform,[4] to capture a dataset that included hyperlinks to our selected profiles and their latest 30 posts.

### 3.2 Data Analysis

We began by importing our collected data into NVivo 14 qualitative analysis software. Drawing upon Ziskin's [147] data analysis strategy within critical discourse analysis (CDA), we wrote reflexive memos for each profile, documenting our initial impressions. Some questions we sought to answer during this memo-writing process included: Is it clear which caste community the profile represents? How do we know the profile voices upper caste interests? Is the profile dedicated to a particular caste community or a particular cause around caste? Does the profile suggest a clear agenda that is related to caste? These memos helped us validate our selection criteria for each profile and establish the necessary context for subsequent analysis.

Using the thematic analysis technique [25], we analyzed each profile's information (header photo, profile picture, username, bio, and any other information the profile listed) and assigned codes to that profile. For instance, a profile's bio stated that it worked toward the revival of Brahmin culture; we coded this profile with the codes "nostalgic about the past," "information control," and "visibility." Next, we coded the latest 30 posts (sorted by time of posting) and the first 5 replies (shown to the study account by X's algorithm) under each post. Our codes especially addressed the rhetorical purpose of the pieces of content (post, image, reply, video) and their corresponding affordances. This coding process helped us translate our data into meaningful categories for further analysis. After the first round of coding the dataset, we iteratively compared the codes with one another and with the data, engaged in discussions among ourselves, wrote memos to document our emerging insights, and revised the codes. We finally merged related codes to generate a set of relevant themes and identified connections between themes.

[4]Meltwater, an online media company founded in Norway in 2001, samples publicly available content from a range of sources, including blogs and news platforms (e.g., BBC, The New York Times), social media sites (e.g., Facebook, X, Reddit, TikTok), and video sharing platforms (e.g., YouTube, Twitch). To learn more, see Meltwater's About Us page: https://www.meltwater.com/en/about

We used the CDA [43] approach throughout our analysis, which helped us focus on how social-power and inequality are enacted, reproduced, legitimized, and resisted by online text and talk [140]. van Dijk [140] writes that one of the tasks of critical discourse analysis is to identify forms of domination. At its core, CDA explicates the various discursive practices that dominant groups use to maintain power. Toward this end, we treated caste-positive accounts as manifestations of caste cultures, spaces, and discourses that, in turn, reproduce the caste structure itself. This meant that during data analysis, our focus was less on the content of the profiles than on the function the content performed. For instance, for a post that criticized a news article about an Indian university's affirmative action policy, we assigned the post with two codes, "anti affirmative-action" and "expressing caste pride" instead of codes specific to the case described in the news article. As mentioned in the introduction of this paper, we focused on the various discursive practices laid out by Wodak [142] and conducted a discourse analysis that interrogated how caste-positive communities construct in-groups and out-groups and simultaneously portray themselves in positive light and lower-caste communities in a negative light. The findings reported below have evolved from our reading of the memos generated, our codes and themes, and regular and in-depth discussions between authors.

Given that the profiles are entirely public and community-driven, i.e., not associated with any one individual, we expect minimum risk to the members who manage them. We decided not to paraphrase the content to completely de-identify the profile. However, we have blurred usernames to minimize the risk of violating the privacy of individuals who engaged with the studied profiles.

### 3.3 A Note on Positionality and Reflexivity

Recent critiques of positionality statements discuss how the performative act of acknowledging privilege does nothing to disrupt power relations between the oppressor and the oppressed communities. Instead, it reinforces it within academia by assuaging the oppressor's guilt [52]. In response, we use this section to go beyond merely acknowledging our upper-caste privilege by more deeply considering upper-caste complicity [5] and the need to study upper-caste cultures.

Both authors were born and raised in India and have witnessed first-hand the role of caste in dictating individuals' financial ability and social capital. Our position as members of upper-caste communities (the first author comes from a Brahmin community in South India and the second author comes from a Vaishya community in Central India) informs our research interests since we believe that examining discourse within upper-caste cultures is vital for understanding the normative status associated with upper-caste identities. Our shared backgrounds with the studied communities helped us understand the sociocultural and religious contexts of our data in a more nuanced fashion. Our understanding that upper-caste complicity often manifests through an active erasure of caste has shaped our inquiry. As upper-caste researchers studying caste, we acknowledge that we are complicit in participating in scholarship that continues to be designed to favor our perspectives as opposed to those of the marginalized, whose voices are underrepresented within academia. But with our research, we intend to not evade but encourage purposeful and critical reflection. We situate our research within one of the strands of critical caste studies [7] and CCTS [121], which is to study the naturalization and normalization of caste. We do so by studying our own contexts or what can be called studying *within*. We affirm that discriminatory cultures are deeply rooted and inhibit change by preventing dominant castes from confronting caste injustice. In response, anti-caste praxis insists that along with empowering the oppressed, transforming the oppressors is just as essential for systemic change and collective liberation. With this paper, we hope to provide a reflexive critique of caste power and contribute to the practice of social justice scholarship.

Drawing upon Cifor and Garcia's [31] idea of "self-disclosure" as a method and a practice in reflexive research, we also want to map our professional identities and experiences as they affect our study of the social category of caste. The two authors' research experiences have proved to be complementary. The first author has experience conducting qualitative research to understand how caste intersects with education [85, 86], governance, and digital infrastructures in urban and rural India [87–90]. Additionally, the first author has worked on documentary projects that have helped her recognize how caste intersects with class and gender. The second author has worked extensively on understanding the perpetration of online harm (e.g., hate speech, online harassment) against marginalized communities [61, 78, 79] and examined how content moderation policies, administrators, algorithms, and designs could address such harm [29, 72–76, 80, 81]. In addition, the second author has conducted digital ethnographies to explore how perpetrators of online harm perceive and justify their actions [77, 79, 141].

Given that this research falls under the interpretivist paradigm [56], particularly social constructionism, our emphasis is on constructed and shared meanings within the discourse. We acknowledge that meanings are not static and our understanding of India's socio-cultural context, our experiences of living in upper-caste spaces, our status as working in the diaspora, and our immersion in prior literature have influenced how we have made sense of the collected data.

## 4 Findings

While revising and synthesizing our codes, we observed that the discursive practices of online caste-positive community profiles are characterized by two broad, complementary narratives contributing to the construction of in-groups and out-groups as well as positive self-presentation and negative other-presentation [139]. One narrative asserts that upper caste cultures deserve to enjoy a superior status in the caste society; the other claims that upper castes are the "true" victims of caste oppression and that there is a need to revive the past that keeps caste inequities intact. In this section, we present the discursive practices that contribute toward these two narratives in terms of rhetorical and organizing strategies as both are concerned with the construction and exchange of meanings within the context of caste. While *organizing strategies* refer to the ways in which members coordinate and structure their activities, roles, and relations in a group, *rhetoric* involves the use of symbolic action and language to influence, persuade, and construct the shared meanings [67]. We critically analyze textual and audio-visual excerpts from our sampled profiles to identify the tactics and affordances promoting this convergence and their influence in reproducing caste. We refer to the overarching "general choices" made by the accounts as "strategies" and the more "specific choices" — guided by these overarching strategies — as tactics [23]. **Table 1**, summarizes our findings to illustrate how caste hierarchy obtains legitimacy in and through online mediation. We discuss the specific instances and observations identified from our dataset by indicating their corresponding profile numbers or P#.

| Strands of Argumentation | Tactics | Rhetorical Strategies | Organizing Strategies | Key Affordances |
|---|---|---|---|---|
| Asserting caste cultures and deeming them to be superior | Building dedicated spaces | Space acting as a rhetoric, eliciting a sense of belonging | Homophily-seeking and driving engagement | Information control, visibility, searchability |
| | Expressing caste superiority through audio-visual cultures | Religious authority, documentary authority, rhetoric of pride | Audio-visual means as resources to build awareness | Bandwidth |
| | Undermining anti-caste measures | Memetic humor, mockery, and hate speech | Facilitate bonds among members based on shared laughter and frustration | Bandwidth, shareability |
| | Consolidating caste networks | Maintenance of caste-purity and caste-loyalty | Forging and strengthening intra-caste networks | Visibility, searchability, shareability |
| Claiming victimhood | Appropriating narratives of the oppressed | Perception of being unfairly treated | Online campaigns | Information control, visibility, searchability |
| | Building evidence in support of their victimhood | Eliciting affective responses of empathy toward upper castes | Sourcing and sharing data, infographics and testimonials | Bandwidth |
| | Portraying constitutionalism as the problem | Eliciting affective responses of anger, frustration, pessimism, and/or desire for change | Demanding policy changes | Bandwidth |
| | The push for revival | Eliciting affective responses of nostalgia for their cultures and the threat of losing them | Digital archiving and policy change efforts to retrieve and restore Hindu upper-caste cultures | Bandwidth, visibility |

Table 1. Table summarizing strands of argumentation with corresponding tactics, rhetorical and organizing strategies, and key affordances.

## 4.1 Asserting Superiority of Caste Cultures

Culture can be interpreted as entangled manifestations of social practices, norms, beliefs, and symbols of particular groups of people, each carrying shared meanings that the group relies on to communicate and form bonds [56]. For caste-positive communities we analyzed on X, this process manifests through upper castes' tactics such as (1) building dedicated spaces online, (2) expressing caste superiority through audio-visual cultures, (3) undermining anti-caste measures, and (4) consolidating caste networks.

*4.1.1 Building dedicated spaces online.* The first tactic involves caste-positive communities carving out spaces on X and beyond to promote upper-caste variations of different cultural aspects, such as religion, rituals, family life, food, and language. This was done using the profile information section to define membership criteria, hashtags to gain traction, and links to usher followers into spaces beyond X.

**Using profile information section to define membership criteria.** Each account on X can manage self-presentation by editing their profile to display their name, a 160-character 'bio,' location, website, cover photo, and profile photo. We observed that caste-positive communities build dedicated spaces online by leveraging this profile information section of their account to clearly indicate the group(s) they are conversing with. For instance, P8's bio stated that it is an "Online community for the #brahmins worldwide #tamilbrahmins." P21 used the bio section to communicate that it is dedicated to "collate historical data of Madhva community [a Brahmin community in southern India]." Instead of using the bio section, P41 used its profile photo to suggest that it "proudly" voiced the interests of the "unreserved" or the "general category" — a political term for those who are not entitled to affirmative action, i.e., the upper castes. By revealing information about their interests and explicitly calling upon members of specific groups, the profiles build virtual roofs that invite in-group members to join in.

**Using hashtags to gain traction.** Several profiles used hashtags regularly in their posts, replies to other accounts, profile information, and images. For example, P12, a Brahmin community account, used its profile name as a hashtag in every post. Upon clicking the hashtag, all posts made by it and its supporters appear in the same place. The account also featured multiple hashtag variations to reflect the different colloquially uttered names of the community, possibly because the community can then link more than one set of search results to their profile and its network.

**Using links to usher followers into spaces beyond X.** Caste-positive profiles mention links to their profiles on other platforms for their followers and other community members to seek out and follow. This practice explicitly recognizes their intention to engage with their followers on multiple social media platforms. For example, several accounts (P13, P17, P15, P20, P21, P24) had hyperlinks to their Instagram profiles, websites, email, WhatsApp groups, YouTube, and Telegram channels in their bios and, in some cases, their pinned posts[5]. Instead of links, some accounts promoted their profiles by putting icons of other platforms on their header photos. This shows that most accounts exhibited significant diversity in their modes of engagement, i.e., their space often extended beyond X.

The plausible outcomes of these tactics are twofold. First, toward organizing goals, these tactics help accounts solicit connections with like-minded people online and drive engagement with their content on X and beyond. Second, such tactics demarcate the profiles as dedicated places for members of the specified caste communities. One might argue that the profiles are public and any person, irrespective of their caste, can view and interact with the account. However, the profiles explicitly and implicitly invite members of their caste community to elicit a feeling of caste-based belonging. Such profile articulations highlight the rhetorical role of online spaces in the definition and maintenance of caste-based identity. The direct invitation makes an online (and publicly visible) space into one that is dedicated to the expansion and strengthening of upper castes' sociality.

*4.1.2 Expressing caste superiority through audio-visual cultures.* The second tactic we observed is that caste-positive communities harness the audio-visual potential of X to produce content that expresses caste pride and superiority. The affordance of bandwidth, in this case, enables profiles to explore diverse and creative modes of caste expressions. This was done by incorporating caste-markers, religious authority, and documentary authority to legitimize caste superiority.

**Incorporating caste markers.** Brahmin community accounts (P7, P11, P47) displayed photographs, videos, and illustrations of men clad in *pooja dhotis* or *sovales* (unstitched garments made of silk that are worn during religious occasions) and the *janeu* (the sacred thread). In many photographs, they are seen performing idol worship (**Figure 1**). The *sovales*, the *janeu*, and the act of touching the idol function as caste markers since they carry immense symbolic power. The *sovales* (worn by men) are believed to develop and sustain greater religious and spiritual affect in the worshipper and his environment, and the *janeu* is worn traditionally across the torso after a boyhood rite of passage ceremony that is exclusively practiced by the upper castes, especially the Brahmins, thereby serving as potent communicative elements of caste.

[5]A post that is saved to the top of one's profile on X, formerly known as a pinned tweet.

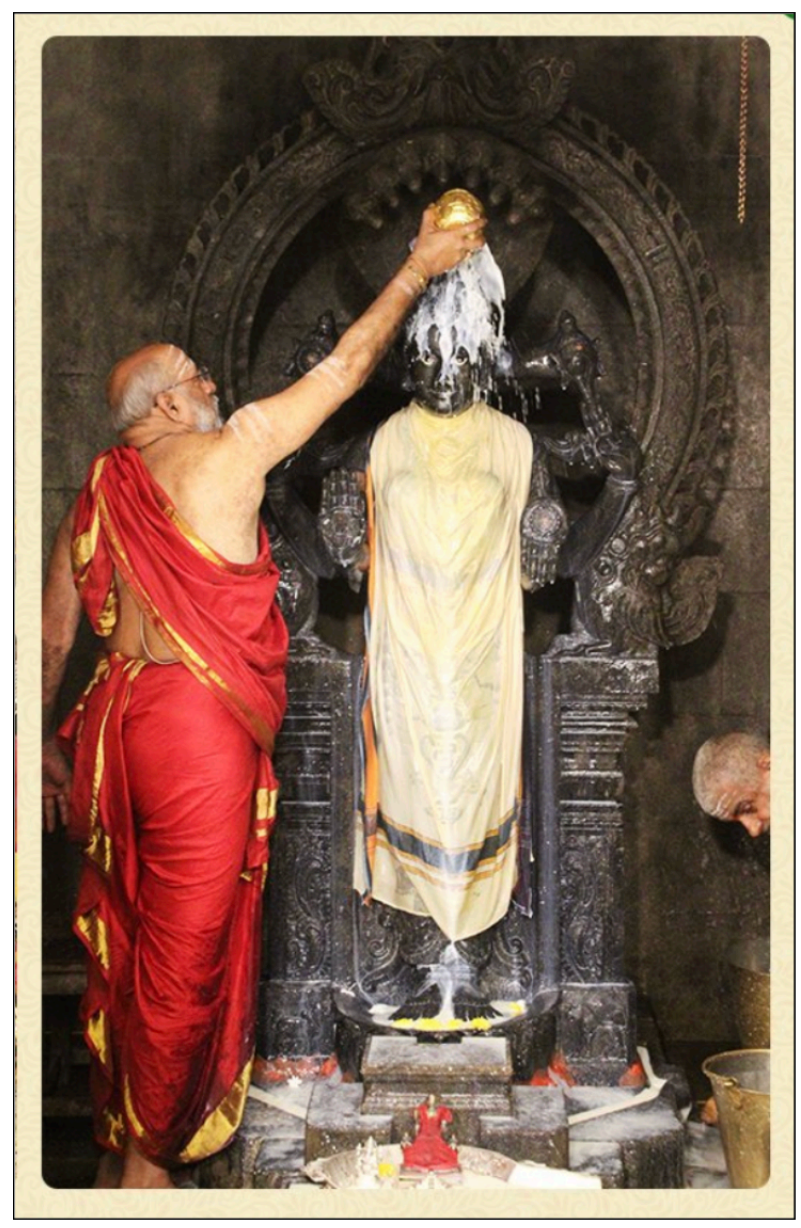

Fig. 1. Image posted by P47 showing a Brahmin priest performing idol worship wearing a janeu and sovale. Source: X.com

**Incorporating religious authority.** Another mechanism by which the accounts express caste superiority and claim the legitimacy of their status is by invoking the rhetoric of religious authority through audio-visual means. Caste hierarchy's obvious intersection with Hinduism is expressed and made clear on the profiles since the accounts regularly call upon deities such as Ram (a Hindu deity who is referred to as a king of the land) and Parshuram (a Hindu deity who is considered to be a Brahmin sage and a warrior) on their profiles through posts and images. In addition, many profiles also produce content in the form of photographs, short videos, video essays, podcasts, and even images generated using artificial intelligence that discuss how the caste hierarchy is ordained by Hindu religious scriptures. For instance, **Figure 2** shows a screenshot from a short video of a man dressed as the deity Parshuram, holding his axe in his right hand and bow in his left. The video, posted by P36, is edited with special effects, and the background music is composed with fast electronic beats against a male voice singing the lyrics *"Brahman hai hum! / Nahin kisi se bhi kam / Brahman hai hum! (Jai Parashuram!),"* which translates to "We are Brahmins! / We are less than no one / We are Brahmins! (Hail Lord Parashuram!)" By invoking the deity Parashuram, the video uses the rhetoric of religious authority to claim its superior status, i.e., ascriptions of their status and traits are associated with a divine purpose that legitimizes their position in the caste hierarchy.

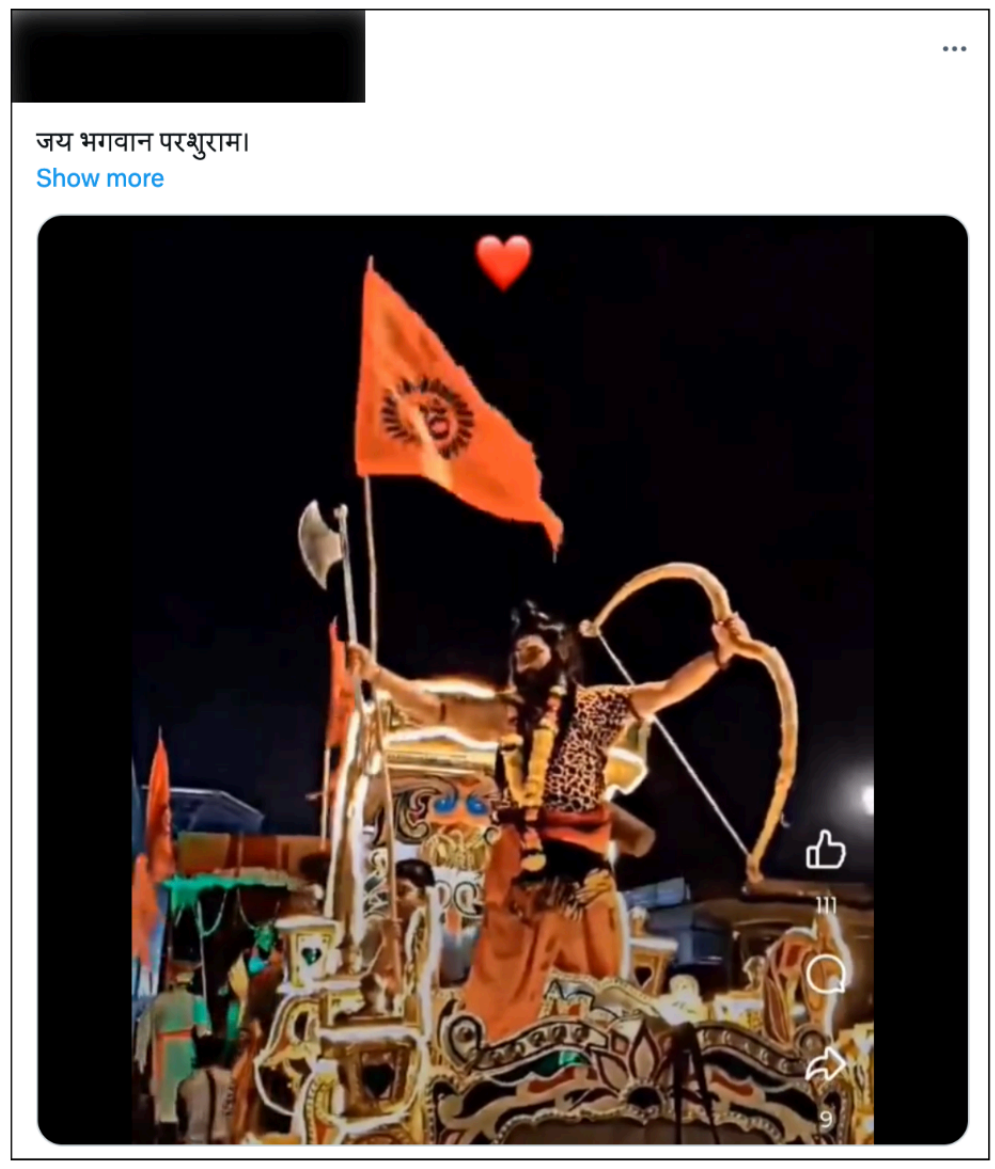

Fig. 2. Screenshot from a short and heavily edited video posted by P36 that expresses caste pride. The man in the image is dressed as Hindu deity Parashuram (a Brahmin sage and warrior). The post says, "Hail Lord Parashuram!" in Hindi. Image posted by P47 showing a Brahmin priest performing idol worship wearing a janeu and sovale. Source: X.com

**Incorporating documentary authority.** Several accounts gathered, processed, exhibited, and quoted from a diverse range of archival texts and artifacts, including newspaper clippings, blogs, photographs, maps, videos, paintings, sculptures, monuments, coins, and stamps. These textual and archival materials were incorporated into attention-eliciting videos or woven into informative threads narrating their communities' stories. This made the profiles act as a visually rich collage capturing the communities' key historical events, influential personalities (such as kings and freedom fighters), and places of significance. While sharing textual documents, some accounts used the ability to highlight digital documents to mark excerpts to direct their readers' attention. For instance, P30 highlighted an excerpt from an unverified archival document that claims that Brahmins are superior. The post's caption says, "Hail the Brahmin Lord," and shows a red flag emoji, a frequently used symbol among Hindu nationalists (see **Figure 3**).

These examples show how a Hindu-centric and caste-driven audio-visual culture continues to evolve on X. By using various media types and archival material and weaving them together with caste markers, the community profiles not only create and circulate a large corpus of audio-visual resources but use them to call upon religious and documentary authority. In addition, the audio-visual materials perform a key rhetorical function of invoking strong affective responses of caste pride among the members. The content contributes to the accounts' organizing efforts by performing a pedagogic function as it attempts to cite Hindu texts to raise the community members' (in this case, the Brahmins') understanding of their so-called superior status.

*4.1.3 Undermining Anti-caste Measures.* Upper-caste cultures represented in these profiles also include discriminatory mechanisms wherein lower-caste communities and anti-caste cultures are severely undermined. Some accounts harnessed the affordance of shareability to circulate potentially viral memes, covert mockery, and hate speech that attempt to diminish and even demonize the lower castes.

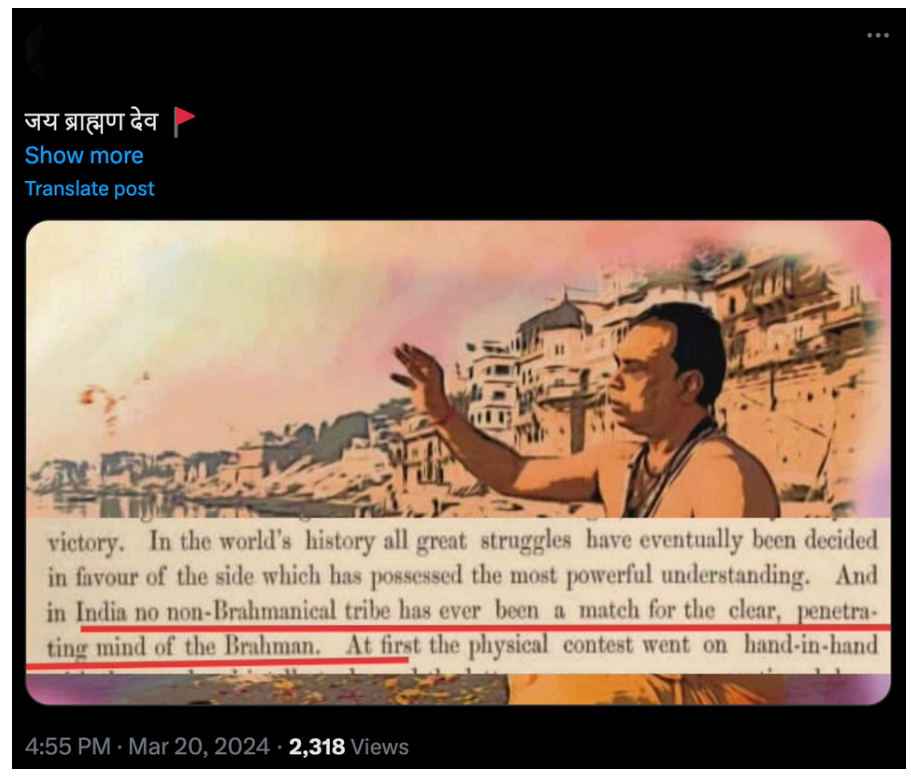

Fig. 3. A post by P30 displays an excerpt from an unverified source claiming that Brahmins are superior to non-Brahmin communities. A man is seen raising his hand while performing pooja (worship) on the banks of a river, with town/city buildings in the background. Source: X.com

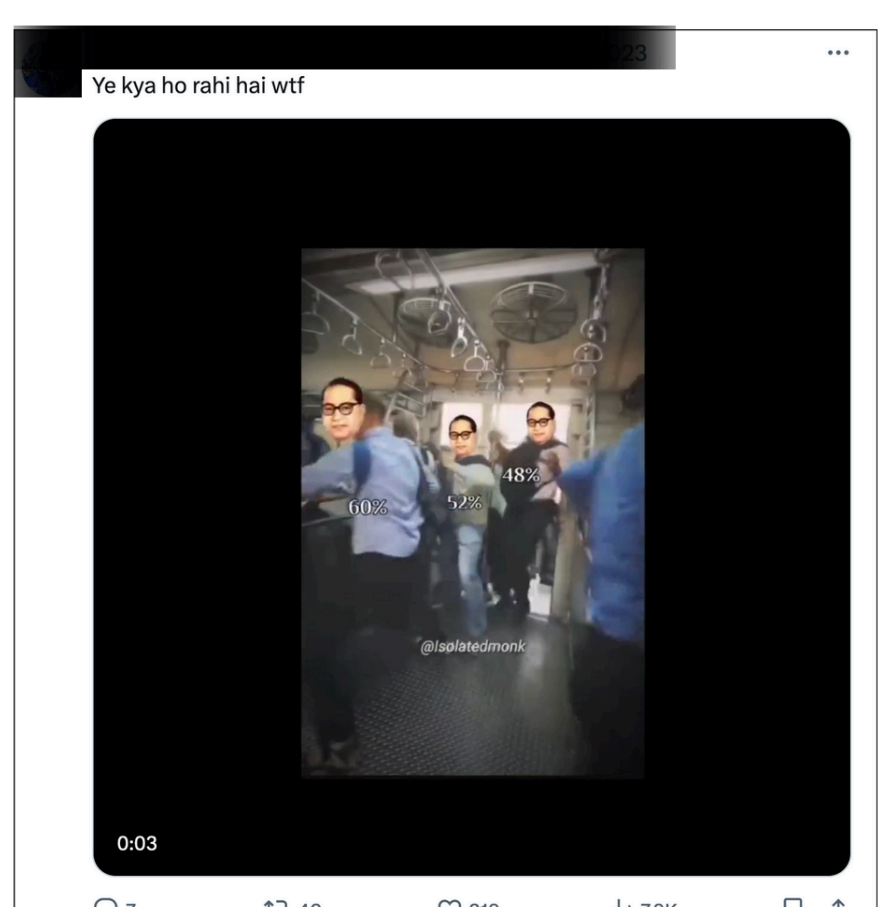

Fig. 4. A meme posted by P39 that mocks the constitution's affirmative action policy by suggesting that a major share of the seats in universities and the government (represented by percentages) is taken by lower castes (represented by people with Ambedkar's face). Source: X.com

**Memes.** Memes were a significant part of the visual rhetoric that the upper castes used to distinguish themselves from the lower castes. Take, for example, the meme posted by P39 in **Figure 4**. It suggests that lower caste communities (represented as people with Ambedkar's face who are seen running into a train compartment) are taking up significant shares of seats in universities and the government with the support of affirmative action. The post's caption exclaims in Hindi, "Ye kya ho rahi hai wtf [sic]," which translates to "What the f*** is happening?" This post's implications are twofold. One, it suggests that the constitution's affirmative action favors the lower castes disproportionately; two, it elicits humor and pride among upper-caste audiences who are meant to feel a sense of superiority due to their “unreserved” status.

**Covert mockery.** Some of the accounts’ content used indirect techniques to mock lower caste communities and their anti-caste struggles. P40, for instance, relies on rhetorical fallacy to elicit such mockery. Most of its posts take valid anti-caste arguments, over-simplify them, and repackage them for an upper-caste audience. On the first read, the profile appears to be concerned with the rights of the lower castes. However, reviewing the profile and interactions in the comments more closely, it becomes clear that the account attempts to mock anti-caste efforts. Several of the comments are just a string of laughing emojis. One comment said, “Seriously this is the best parody account on Twitter” followed by clapping emojis. It becomes apparent that this parody account is masking casteist remarks when we read comments such as this:

> “Sarcasm aisa kro ki log confuse ho jayein ki real h ya humour. (Display your sarcasm in such a way that people get tricked into wondering if it is real or just for humor.)”

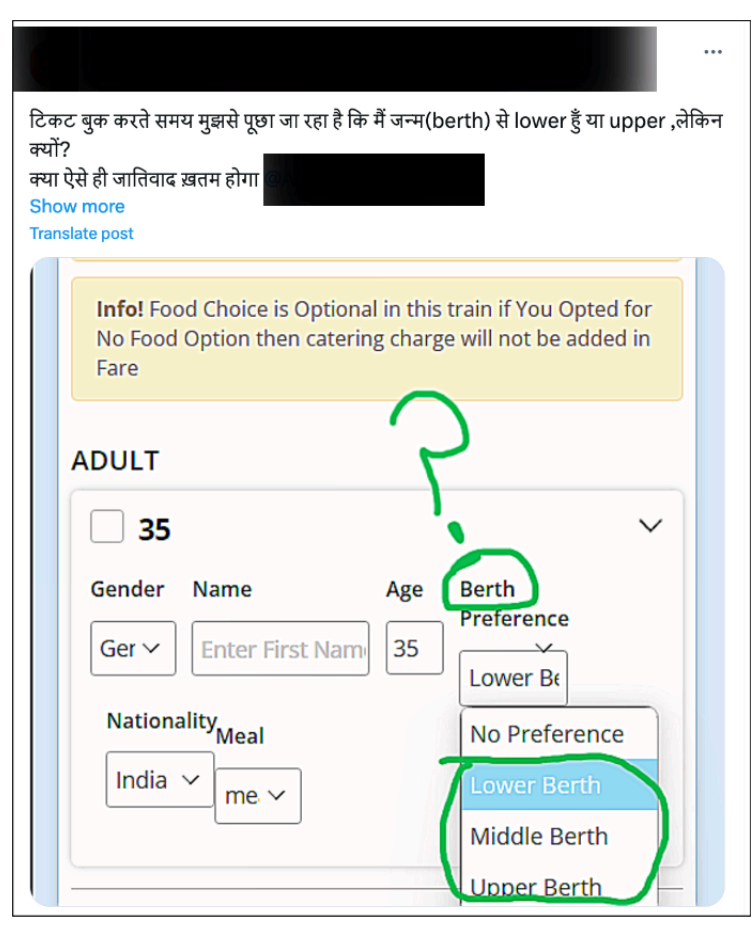

Fig. 5. P40’s post that displays a screenshot from a train booking website. The post intentionally confuses the word berth with birth and asks, “Why am I asked to select if I prefer lower berth [or birth as a lower caste] or upper? Casteism cannot be destroyed if this continues.” By doing so, the post exaggerates, mocks, and attempts to undermine critiques of caste. Source: X.com

The profile grossly exaggerates the anti-caste critique and mocks it in ways that undermine the original argument’s credibility. For instance, **Figure 5** is a post by P40 that shows a screenshot of a train booking website. It intentionally confuses the word *birth* with *berth* and asks, “Why am I asked to select if I prefer lower berth [or birth as a lower caste] or upper? Casteism cannot be destroyed if this continues.” By over-simplifying the critique of the caste system, taking it out of context, and presenting it as if it were sincere, this post attempts to reduce the struggle for the eradication of caste to entertainment.

**Hate Speech.** Accounts also resort to discourse practices wherein normal language or spellings are twisted to create obscure phrases and symbols that denigrate lower castes. These phrases and their hidden meanings eventually become common knowledge among the members. For example, we found that derogatory words “Bhimta” and “Dumbedkarite” (Dumb + Ambedkarite) are used to refer to lower-caste communities, especially those who align with Ambedkar’s (whose first name is Bhimrao) politics. To avoid being detected by anti-caste groups or flagged for hate speech on the platform, P39 replaced “i” from the term with an exclamation mark (**Figure 6**). The post also uses an emoji (a hand gesture) to suggest that lower castes have low Intelligence Quotients (IQs). To criticize the anti-caste politics of people in the South Indian state of Tamil Nadu without attracting their attention, the account refers to the people of the state as “Dumeel” (Dravidian + Tamil). Given the obscure and situated nature of such hate speech, the content posted by these accounts often evades scrutiny and moderation.

In sum, the process of marginalization on X encompasses various modes of undermining the lower castes, including memetic humor, mockery, and hate speech. Such rhetoric evokes a positive self-image by presenting a negative image of the other. We posit that beneath the humor and casual mockery lies the reactionary resentment that some upper castes hold against lower castes and their struggles and successes in achieving equal opportunities.

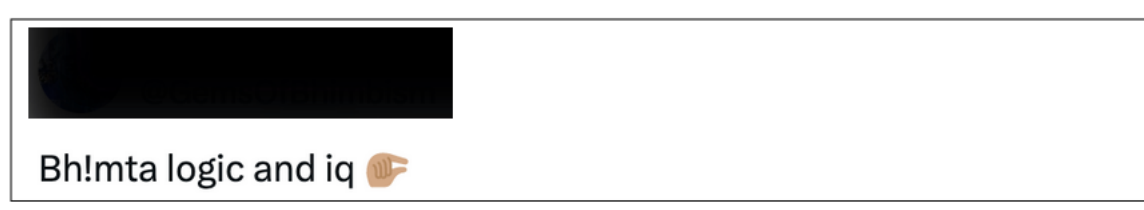

Fig. 6. P39’s post containing caste-based hate speech that uses discrete methods (special characters and emojis). Source: X.com

*4.1.4 Consolidating Caste Networks.* The fourth tactic that we observed was building caste-based affiliations and translating the network into caste capital. This organizing of capital manifests in the form of online campaigns, extends profiles’ offline activities, and garners support for their causes. This tactic primarily uses the affordances of visibility, searchability, and shareability to build and sustain networks.

**Organizing online campaigns.** Caste-positive accounts mobilize support for their causes by deploying features such as hashtags and tagging to gain the attention of their community members, influential individuals (such as government ministers), and key public institutions such as the judiciary, the press, and universities. As part of such campaigns, we noted that one key goal is to “mass-post” using a hashtag until it begins trending on X. For instance, P10, a caste-positive local media organization, celebrated the trending of the hashtag #हम_मनुस्मृति_पूजेंगे, which literally translates to “we will worship the Manusmriti,” a Hindu religious text that is claimed to be the oldest jurisprudence and is heavily criticized for codifying the caste system. This hashtag trended on December 25th, the Manusmriti Dahan Divas (the day on which Ambedkar publicly burnt the text as an act of dissent in 1927). In response to the anti-caste mobilization against the text (#manusmritidahandivas), the upper-caste communities counter campaigned in favor of the Manusmriti and made the hashtag trend (**Figure 7**).

**Extending offline activities online.** Caste-positive communities use their profiles as an avenue to expand their offline presence and activities into the online realm. This is done by posting photos, videos, invitations, and blogs about cultural events, political rallies, study circles, and other activities that occur in person. For instance, when P12 organized an event on International Women’s Day, an invitation was circulated with the event’s details and a title saying, “Celebrating the achievements of Brahmin Women.” This invitation was followed by posts that acted as reminders until the event day. After the event, updates and pictures from the event were posted, as well. This shows that the community profiles are integrated into the full lifecycle of organizing efforts, wherein they play a key role in initiating the event, mobilizing for it, and documenting it.

**Garnering support for causes.** Apart from organizing events, several accounts used their profiles to build a network wherein the account, its followers, and other members could seek and offer support in different ways. Using the hashtag and tagging features, profiles regularly sought donations (P2) and various materials needed for the community to continue its activities. P21 and P24 sought leads to certain manuscripts that could be added to their digital library. P49 regularly promoted the Instagram and YouTube handles of content creators and influencers from their caste communities to boost their visibility. P9 and P12 also sourced, collated, and shared resources such as job opportunities for their community members' perusal. Yet another example is of P46, which offered "[m]atrimonial service for single, divorced, separated, Disabled Gujarati [people speaking the Gujarati language or living in the western state of Gujarat] Brahmins."

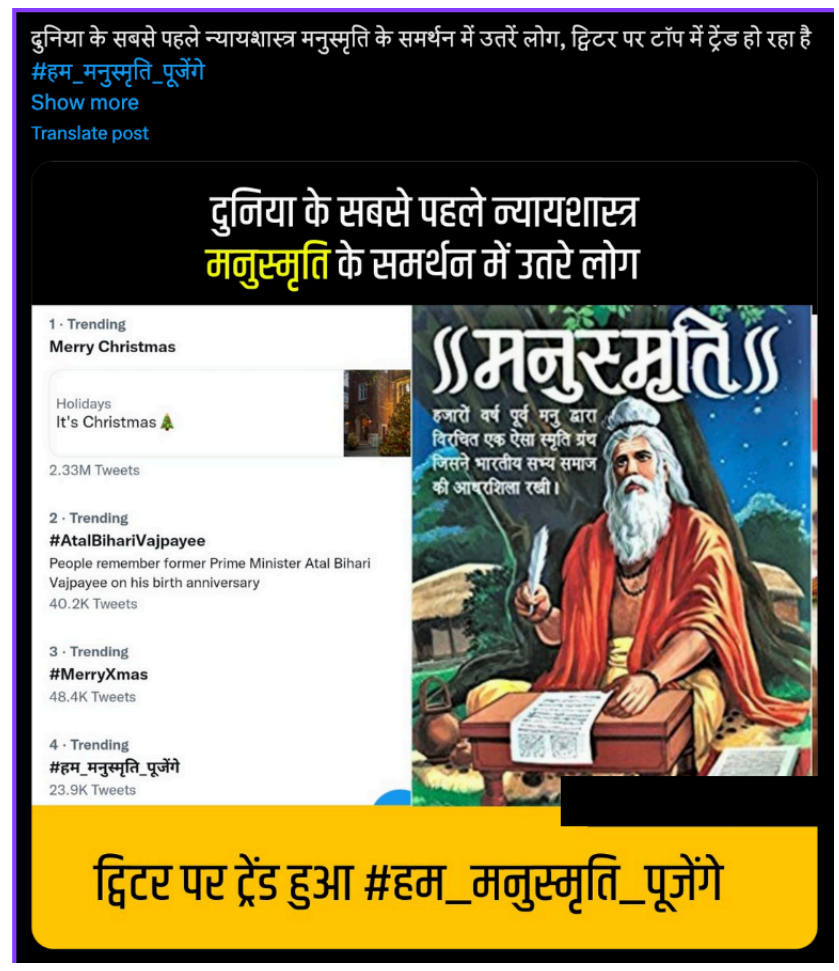

Fig. 7. A post that shows a screenshot of #HumManusmritiPoojenge ('we will worship the Manusmriti') trending. Beside the screenshot is a picture of Sage Manu, the author of the text (P10). Source: X.com

Such organizing activities (campaigns, event updates, promotions, and matchmaking services) aim to forge tight intra-caste networks that often transcend geographical boundaries. They also double as rhetorical strategies since they prioritize and elicit powerful emotions of caste-loyalty and caste-purity. The establishment of a virtual network that allows exchange of resources, favors, and support ensures the consolidation of caste capital that can be shared among the members of the community.

## 4.2 Claiming Victimhood

The second argument strand that upper castes use to circulate and reinforce caste ideologies maintains that they are the "real" victims. This was expressed through various tactics, such as (1) appropriating narratives of the oppressed groups, (2) participating in evidence-making processes, (3) blaming India's constitutionalism, and (4) insisting on reviving India's "past" and "culture."

*4.2.1 Appropriating the Language of Resistance Movements.* Despite upper-castes holding relatively more social, cultural, and economic capital [21], they create profiles that tend to portray themselves as truly marginalized communities. Toward this end, the profiles incorporated techniques of progressive movements into their profile information, hashtag usage, and content to indicate their so-called victim status.

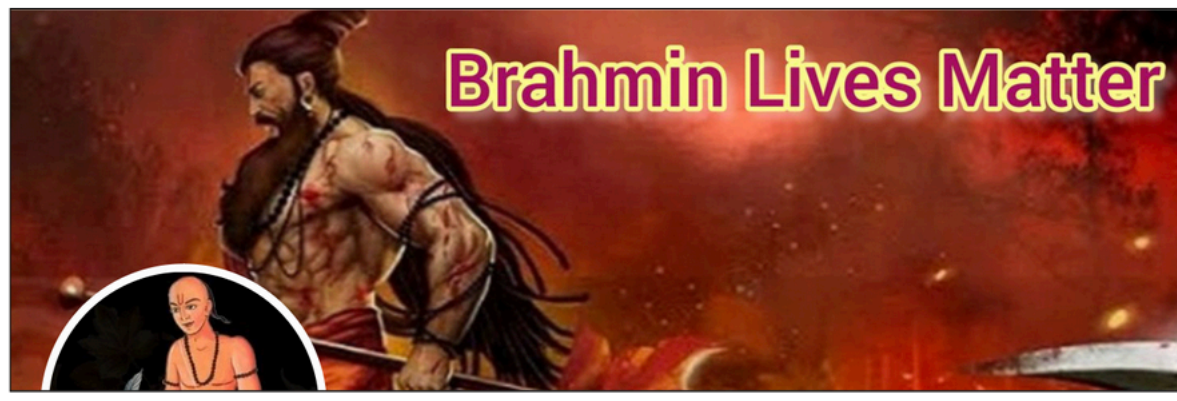

Fig. 8. P35's header photo displaying the text 'Brahmin Lives Matter.' Lord Parshuram is seen walking aggressively covered in blood and carrying his axe (believed to be his weapon). Source: X.com

**Profile information.** P35's header photo (**Figure 8**) reads "Brahmin Lives Matter" in bold letters. The text is accompanied by Parshuram walking aggressively, covered in blood, and carrying an axe. His posture in the image and the burning trees in the backdrop suggest anger and fury against the oppression Brahmins claim to face because of their caste. Another example is that of P14, whose bio reads, "SLM is Sawarn-Rights movement Since 2005 to abolish Prejudice,Bias against Sawarn [sic]." On many profiles, the textual and audio-visual content available in the profile information section had a diverse range of communicative elements to co-opt imagery, slogans, and symbols used by the oppressed.

**Hashtags**. Profiles often appropriated and adapted hashtags of resistance movements to suit their purpose, including facilitating online campaigns. For instance, *#Brahminlivesmatter* is regularly used as a hashtag to attract the attention of others, especially their community members. We identified several other variations of this hashtag, including *#SavarnaLivesMatter*, *#HinduLivesMatter*, *#Janeucide* (genocide of those who wear the *janeu*), *#MurderofEquality*, *#BrahminsUnderAttack*, and *#MurderofMerit*. Such hashtags enable accounts and their community members to tap into a wider network, mobilize support for their cause, and populate the online spaces in ways that curb anti-caste dissent, specifically, to the call that 'Dalit Lives Matter.' Further, these hashtags create a false equivalence by projecting themselves as comparable to the oppression that Black communities face due to racism [36, 94].

**Content.** Drawing upon the language of anti-caste critiques to legitimize their perception of being oppressed, the profiles often posted content combining various media types. For instance, P10, a Kshatriya community account, posted a poem claiming that the political institutions, media houses, and cultural industries do not belong to them, and this lack of representation and ownership causes their issues to be neglected. Even though the Kshatriya community enjoys an upper-caste status, is politically powerful, and is historically a landowning caste in northern India, the account portrayed its community members to be the victims. This tactic depicts what Anderson [3] calls "linguistic hijacking," wherein the oppressor communities build their victimhood narrative by co-opting expressions, language, and aesthetics that are "epistemically relevant to the pursuit of social justice."

The same platform affordances that the profiles use to assert the legitimacy of their higher position in the caste hierarchy — visibility, information control, searchability, shareability, and bandwidth — are also used to construct a self-image of being oppressed and seeking affirmations from community members across all privileged castes. This is evident from a reply under P10's post, which said, "Full support *sahb (*I am with you, sir*)*. Being [a] brahmin boy, I can feel it. bcoz we are also being targeted by them for caste discrimination and all." Such narratives facilitate the consolidation of upper castes via a shared affect of being unfairly treated — yet another instance wherein rhetorical and organizing efforts act in concert.

*4.2.2 Building Evidence in Support of Their Claim to Victimhood.* We observed that the caste-positive community profiles' contention that they are marginalized was accompanied by efforts to build evidence in its support. To convince their audiences, the profiles contributed to building a media ecosystem, circulating statistical data, and sharing testimonies.

**Building a media ecosystem**. For caste-positive accounts, this tactic involved curating profiles to serve as news sources for their audiences. For instance, P43's bio said, "Fastest Growing Digital Hindi News | Follow For Latest News & Updates | Voice Of General Category." The phrases "Fastest Growing Digital Hindi News" and "Voice of General Category" attempt to establish the media house's credibility, relevance, and trustworthiness. "Follow For Latest News & Updates" is a call to action that increases its following. In terms of content, the profiles were invested in serving as news sources for their audiences. This involved documenting crimes against upper castes and portraying them as caste-motivated despite lack of verified evidence (P28, P35), campaigning in favor of political candidates belonging to their caste community (P3, P4, P12, P34, P48), and rallying against anti-caste political parties (P39, P43).

**Circulating statistical data**. The profiles sourced and shared quantitative data and infographics to argue that their groups are the society's "real" neglected ones. For instance, **Figure 9** is P13's profile picture, which suggests that the populations that seek reservations, i.e., the marginalized castes, occupy a majority of the seats in universities and government positions, whereas the unreserved or the "meritorious" are given a lesser share. This infographic does not specify any numbers or sources but is used merely as a rhetorical device to construct an Us vs Them narrative. In fact, most of the data points that the profiles shared mentioned no source, making it difficult to verify them. Some data were evidently fabricated. For example, P29 posted fake data to claim that America's healthcare system is better than India's because 50% of its workforce are Brahmin doctors from India. This was accompanied by #Justic[e]4General, a common hashtag used to advocate upper caste interests.

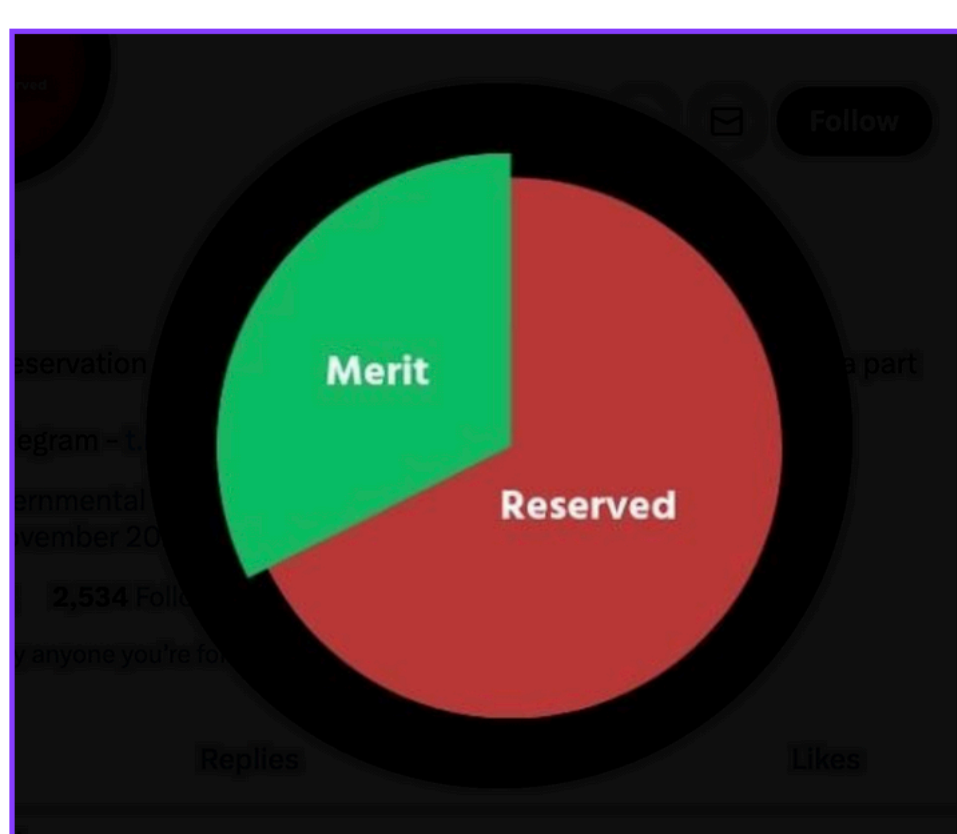

Fig. 9. P13's profile picture is a pie chart showing that the reserved populations (lower castes) have the majority. The pie chart provides no detail, data or source. Source: X.com

**Sourcing and sharing testimonies.** Some profiles invested in sourcing and circulating testimonies from community members about instances of unfair treatment. This kind of evidence incorporated various media types, including text, images, and videos (P32, P42). In terms of topics, testimonies shared stories of being denied admission at a university or being falsely accused of discriminating against a lower caste member. Such testimonial data is then leveraged to advocate for the upper castes and favor their claim of victimhood status.

Most of these posts have an alarming tone that is meant to invoke empathy toward the upper-caste communities and anger and frustration against structures that, according to them, have enabled their oppression. This tactic of evidence-making, despite its many flaws, serves multiple purposes. It amplifies the narrative that upper castes are the true victims, triggers affective responses of sympathy and anger among the members, and functions as a communication anchor around which the communities can organize.

*4.2.3 Portraying Constitutionalism as the Problem.* We observed that caste-positive communities curated their profiles to undermine the successes of constitutionalism and express resentment toward the lower castes. This is achieved by equating the upper castes with the notion of merit, calling upon nationalist rhetoric of saving the nation's merit, and blaming the lower castes and their anti-caste measures for upper castes' perceived marginalization.

**Equating the upper castes with the notion of merit**. While objecting to affirmative action, several community accounts ascribe merit only to the upper castes and argue that the policy treats them unfairly. This involves representing their community as superior and the lower castes as "undeserving." For instance, **Figure 10** shows a widely shared meme that carries not only caste-positive but racist connotations wherein the Black player (who was also the captain) of the South African cricket team is paralleled with lower castes in India, and both groups are deemed "undeserving." On the other hand, the white players (paralleled with the general category) are portrayed as hard-working achievers (P39).

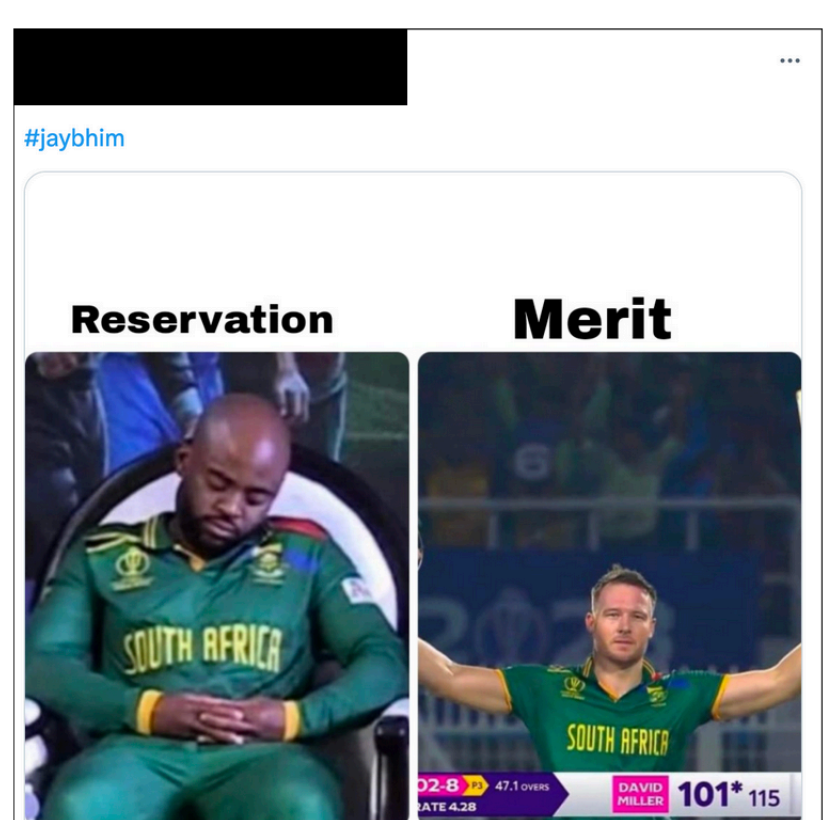

Fig. 10. A meme posted by P39 mocking India's reservation system with the text #Jaybhim (a salutation and a slogan among anti-caste communities). Source: X.com

**Calling upon the nationalist rhetoric.** One of the profiles' (P42) header photo said, "SAVE MERIT. SAVE THE NATION." in bold font. On the left of the text was an illustration of a fist holding a pencil, and on the right was the tricolor flag of India. The photo's extreme left and right portions listed several contacts of legal advisors, lawyers, and activists engaged in gathering testimonies and supporting students who have been denied an opportunity. P38's profile picture reflected a similar sentiment — a pair of hands was shown to protect India's education and vocational training. These communicative and networking elements construct a nationalist rhetoric that argues that upper castes or the general category represent the entire country's merit. Therefore, the argument implies that the policies resulting from India's commitment to equality and justice wrong those who are its merit.

**Portraying the historically oppressed as no longer oppressed**. The profiles posted content to question and delegitimize the oppression of lower castes. For example, P38 posted a family picture of a politician belonging to a lower-caste community. Its caption said: "Family Picture of an Oppressed Backward Caste…[the woman in the family photo] is wearing a Gucci Dress Worth 6 Lakhs." Such phrasing assumes and generalizes about the oppressed castes' wealth and implies that this economic prosperity protects lower castes from sociocultural marginalization. Also implied in the post is the resentment against rich lower castes, as if they ought not to be earning wealth. The accounts also argued that the oppressed castes are the new oppressors. This, for instance, was evident in a post by P38 that referred to Bhimrao Ambedkar — who headed the Constitution's drafting committee and insisted on affirmative action — as "Bhimurai [play on the word Samurai] slicing dreams of the GC [General Category]."

The rhetorical effect of this tactic contributes to a narrative that blames progressive policies and anti-caste efforts for the perceived oppression of upper castes. This results in affective responses that range from anger toward the country and the lower castes, frustration, pessimism, and a desire to bring a collective change. One profile pinned a post that exclaimed, "Hopeless Country and Fake Social Schemes." A reply under the post read, "Reservation should be scrapped, man." Another comment expressed the frustration by saying, "This country is hopeless. Get out of this country as fast u can [sic]. For the future of our children and generations." The organizing attempts ranged from collaboratively deliberating on ideas to proposing possible solutions. To do so, profiles facilitated dialogue in the comments sections, mobilized through hashtag campaigns, and tagged influential people to bring attention to their perceived problems and demand policy changes such as scrapping reservations and banning "freebies" for the lower castes.

*4.2.4 The Push for Revival.* We found that caste-positive profiles' perception of being oppressed and their feelings of anger, resentment, and frustration are accompanied by a sense of threat of losing their status and position in the social hierarchy. In reaction to this, we observed that the profiles seek to save or preserve their culture. This is verbalized by expressing nostalgia toward their "lost" culture, rejecting (and opposing) progressive changes in society, and attempting to restore caste-coded Hindu cultures.

**Expressing nostalgia toward their "lost" culture.** Reviewing the accounts and their descriptions of their culture suggested that they long for a kind of nation where caste norms and rules were more strictly enforced. This is evident from P2's post, which shared a documentary about rituals performed primarily by Brahmins. The post mourned the loss of such culture and wished to see it integrated into society again. We observed a recurring usage of words such as "reviving" and "preserving" in the context of terms such as "culture," "tradition," "language," and "education." For instance, P2 frequently called upon the followers to "come together to preserve the timeless wisdom of the Vedas!" (Hindu religious texts).

**Rejecting progressive changes in society.** The accounts explicitly rejected and opposed the progressive changes that were made possible by resistance movements such as the anti-caste and feminist movements. P15, for example, promoted early marriages of girls and boys and opposed the idea of choice in marriage or even the "boyfriend-girlfriend" culture. The profile argued that early marriage is imperative to prevent "shameful" acts such as pre-marital sex and inter-caste marriages. Instead, it insisted that to protect Hindu culture and the institution of family, it is critical that girls be married off before they engage in relationships outside their caste. In the pursuit of persuasion, the profile targeted their messaging toward parents and argued that they must ensure their children do not transgress social boundaries that their ancestors have established.

This messaging is communicated using various media types, such as text, posters, videos, articles, and photographs. These media objects were supported with hashtags such as #SupportEarlyMarriage, #BanICM (ICM stands for inter-caste marriage), #SayNoToGfBfCulture (gf/girlfriend and bf/boyfriend), and #परिवार_बचाओ (which translates to "save the family"). The account's resistance to social change is also evident in the hashtag #बेटी_ब्याहो_बहू_पढ़ाओ (marry off your daughter, educate your daughter-in-law, i.e., after she is married), a twist on the popular *Beti Bachao*, *Beti Padhao* (save daughters, educate daughters) campaign launched by the Indian government that condemns female feticide and encourages girl child education.

**Attempting to restore caste-coded Hindu cultures.** The accounts translated these preservation ideas into actionable steps by mobilizing support for their causes. For instance, several profiles implemented digital archiving practices to preserve old historical and religious scriptures that often legitimize the caste system. **Figure 11** shows how P2 regularly "retrieves" scholarly texts through digital archival practices. Another actionable step the accounts take is launching online and offline campaigns to bring about policy change. For example, in support of early marriage, P15 continues to campaign to lower the country's legal marriageable age (currently, it is 18 years for women and 21 years for men).

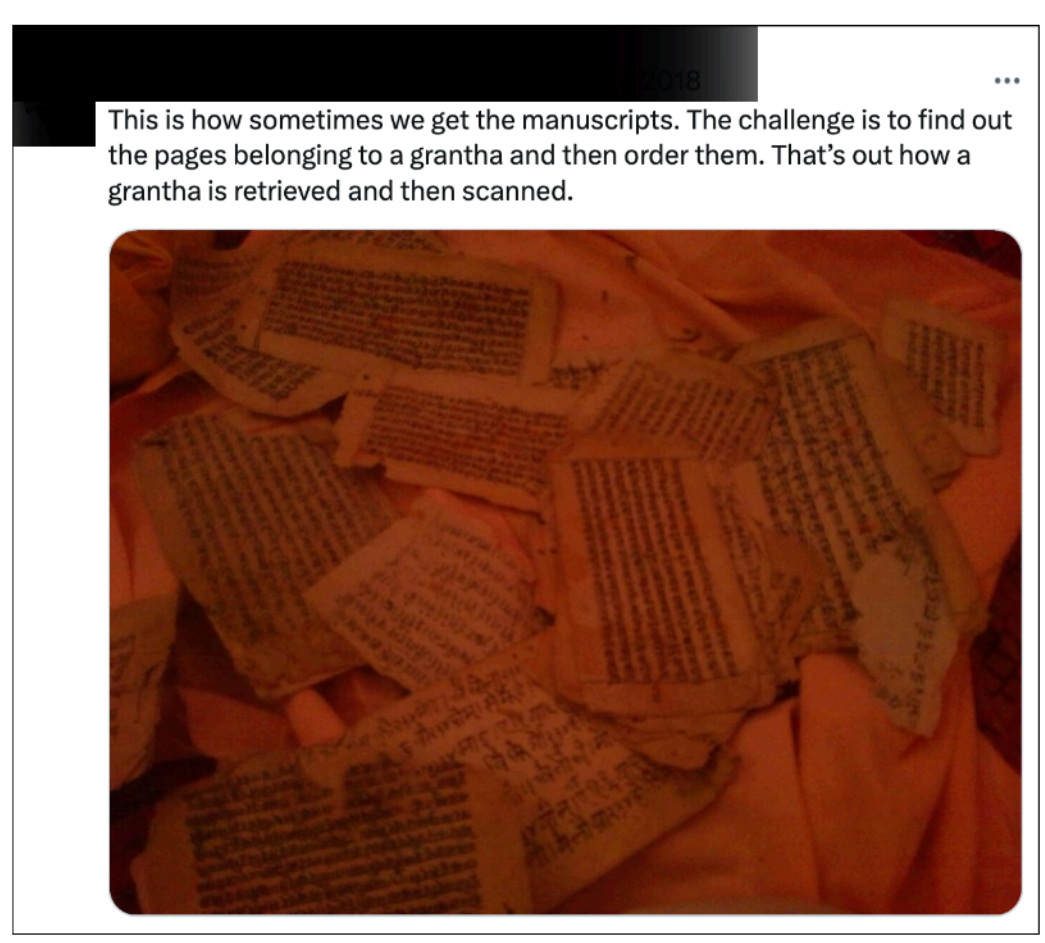

Fig. 11. P2's post sharing how the community sorts and digitizes religious manuscripts; 'Grantha' literally means a scholarly text or treatise. Source: X.com

Thus, the nostalgia for a "lost culture" and the fear of possible transgressions are powerful affective pulls that reinforce the call for a "revival" of the inter-connected oppressive ideologies of caste and patriarchy – together known as Brahmanical patriarchy [26]. The rhetoric of revival implies a desire to achieve a Hindu nation maintained by rigid gender- and caste-based boundaries that, according to them, are fading away. The Brahmin victimhood (and, therefore, the need for a revival) is framed within Hindu victimhood. This tactic of demanding revivalism invariably involves expressing feelings of caste pride and religious authority, thus feeding back into the strategy of circulating caste cultures and asserting caste superiority (Section 4.1).

## 5 Discussion

This paper aims to deepen our understanding of how caste-positive communities communicate and reinforce caste ideologies on X. Our analyses in the previous sections explored how the accounts leverage platform affordances (concerning RQ1) such as information control, visibility, searchability, shareability, and bandwidth toward various rhetorical and organizing strategies that aim to: (1) promote caste cultures and deem them superior (Section 4.1) and (2) claim they are the victims, and their cultures need revival (Section 4.2)(concerning RQ2). We now discuss the implications of this research for the HCI and CSCW communities and scholars who investigate cultural power online.

### 5.1 Constructing Caste-positive Cultures, Spaces, and Discourses

When examining the mechanics of caste, Ambedkar [2] confronted a question not dissimilar from what we began our paper with — what keeps caste prevalent? Ambedkar writes that to sustain exclusionary practices, there is a need to create philosophies around the caste hierarchy that can be normalized among and revered by people. Our paper speaks to this insight by capturing how caste is revered and preserved digitally.

Specifically, in a social-technical environment such as social media, caste-positive communities seek legitimacy using a diverse range of rhetorical and organizing strategies. We found that these two kinds of strategies show significant convergence, that is, they maintain caste while being intertwined. Moreover, they are almost always discursive in nature; specifically, they portray a positive image of the upper-caste communities while simultaneously painting a negative picture of the lower-caste ones. Put differently, the caste-positive community profiles make discursive arguments by constructing worldviews that influence how caste-specific structural power is distributed.

These discursive practices are driven by three key concepts that we discussed in Section 2: culture [2, 22], space [59], and discourses [10, 122]. In the online realm, caste-positive communities' efforts extend these spheres of influence in diverse and interconnected ways. For instance, the accounts contribute to caste *discourses* by initiating dialogues about caste interests through online campaigns, conversations, and memes; evoking and managing the affects of pride, nostalgia, and anger; engaging in knowledge production through sympathetic videos and threads, news, and curation of digital archives. Further, their profiles, the hashtags they participate in, and the trending section of the platform act as dedicated, flexible, and dynamic sites that enable resource- and network-building exercises. Such virtual *spaces* are made to "echo" [95] upper-caste messaging. This messaging also involves the marginalization of the lower castes in ways that mark them and their *cultures* as not only different but in opposition. Accounts execute such representation by undermining lower caste cultures and their successes through memes and "co-opting the position of the marginalized" [40].

Such an analysis is useful for the CSCW and HCI community as our findings suggest that, in the construction of caste, the different affordances play a key role in digitally "interpellating" [1] caste identities. Information control, visibility, and bandwidth enable caste-positive accounts to create textual and media content that articulate caste ideologies. Searchability, on the other hand, enables users to find their accounts online, while shareability allows them to engage with and (re)share content. Affordances, thus, can be looked at as windows into digitally mediated caste articulations. Such an understanding of affordances can be helpful in identifying and characterizing how digital mediations establish "symbolic boundaries" [93] that create and maintain cultural, institutional, and social differences between the upper caste and the lower caste. We posit recognition of these social media strategies in the Global South context as the first critical step toward design, policy, social, and educational efforts that may meaningfully address the systematic harm they cause.

### 5.2 Studying Communities of Power

Our study highlights the importance of critically examining powerful communities in sociotechnical environments such as social media. In particular, one of the key takeaways about power itself is that it is maintained by capitalizing on culture, spaces, and discourses online. Our analysis shows how upper caste communities harness the networking and expressive opportunities of X to deploy rhetorical and organizing strategies to legitimize and sustain their powerful position within the caste hierarchy. More specifically, this is done by incorporating positive self-presentation and negative other-representation in their online activities and interactions. Put differently, our study helps us understand how privileged castes curate their online presence to morally justify [99] their *power and the corresponding marginalization* of the lower castes.

We note stark similarities between the many rhetorical and organizing strategies observed among dominant caste communities and those of the extremist groups in the Global North, e.g., memetic humor [33], use of religious authority and nostalgia for lost cultures [51], engagement in networked harassment to signal group membership [99], and online campaigns [55]. Comparative analyses of these groups' operations should further sharpen our conceptual understanding of online extremist strategies to maintain power. At the same time, it is vital that we move beyond U.S.-based platforms like X and attend to caste dynamics on platforms based in the Global South (e.g., platforms such as Moj and ShareChat), where enacting local policy solutions could be more tenable. We must also attend to how other communities of power in the Global South, such as Hindu nationalists, operate online and examine their connections to caste-positive communities.

Toward advancing scholarship within the field of humanistic and critical HCI [18, 49, 70, 117], our study provides theoretical and methodological suggestions for research invested in studying the mechanics of power along the lines of other social categories, such as gender, race, and sexuality. For one, we frame online discourses, spaces, and cultures of powerful communities as our objects of analysis. These key concepts help us focus on how communities understand themselves as well as other communities. In addition, it does not limit us to only the semantic or linguistic meanings of texts but enables us to pay attention to how power and marginalization are embedded in discursive interactions online. Second, a critical and constructionist approach helps us understand how the diverse social media objects — such as text, images, audio, video, hashtags, and hyperlinks — carry symbolic meaning that moves through a digitally mediated and networked environment. Third, our study's sampling technique can be adapted to gather digital assemblages in ways that isolate discourses of such powerful communities for a focused examination.

### 5.3 Moving Toward an Expansive Understanding of Online Harm

As discussed in Sections 2.2, 2.4, and 4, powerful communities leverage social media platforms to engage with like-minded people and promote their communities' interests. Our findings and the literature review conducted by Kakavand [83] note that their presence on multiple platforms helps them connect with their community members and further spread their exclusionary ideologies, such as caste. Our analysis showed that technological affordances such as visibility, bandwidth, searchability, and shareability were critical in making caste-positive communities' tactics more visible, vibrant, and scalable. What played perhaps an even stronger role than affordances is the accounts' "ability to make their rhetoric palatable" [40] and their organizing effective. This supports existing literature, which notes that online harm is not a straightforward concern but a deep-rooted societal and political issue that cannot be addressed by mere technological fixes or solely focusing on and punishing the "bad actors" [115, 146].

As the field of HCI and social computing confronts this issue, the ways forward for researchers and practitioners are twofold. First, it is helpful to use Scheuerman et al.'s [116] framework of severity and deploy the field's research energies in providing "holistic accounts of harm that can directly inform how to focus interventions into said harm," especially from regions other than the Global North. However, it is crucial to note that the Global South does not exist as a homogeneous extension of the Global North, nor does it exist merely in opposition to it. As our findings demonstrate, countries such as India are textured by intra-regional and inter-regional tensions along the lines of gender, caste, class, religion, and geography that must be accounted for as we expand the meanings of online harm. As Shahid and Vashistha [120] have shown, Western social media platforms reflect coloniality in their content moderation practices when governing content from the Global South, i.e., they center Western norms, peripheralize the needs of marginalized communities, and perpetuate historical injustices. As we continue emphasizing the need for decolonial mechanisms to address online harm, we must simultaneously reflect on whether those mechanisms are also anti-caste. Do they center locally articulated social relations and power dynamics (here, along the lines of caste), especially in countries of the Majority World? **How can our understanding of online harm incorporate justice that goes beyond mere fair treatment and inclusion? How do we include marginalization in our understanding of online harm?** Attending to such questions is vital when extending current conceptualizations of online harm to incorporate caste-based disparities.

Second, it is critical to collectively deliberate on how to move beyond case-based, punitive content moderation and employ more expansive approaches to address online harm. Moderation efforts to sanction individual instances of caste-based hate speech can be helpful [110], but such efforts are far from sufficient. Instead, alternative social-justice oriented mechanisms offer a more promising direction to combat caste discrimination online. Bellini et al. [17] propose the idea of "mosaics of justice" to argue that the field of HCI requires carrying out justice-oriented work along different dimensions, including knowledge production (citational justice and research justice), inclusive design (disability justice), and repairing the "harm(s) caused by actions, behaviors and practices" of the offenders and emphasizing restoration as opposed to retribution (restorative justice). For instance, Xiao et al. [145, 146] note how the restorative justice framework "addresses harm differently than more common punitive models. The main tool for action in a punitive justice model, as embodied in content moderation, is punishing the rule violator. In contrast, in restorative justice it is communication among the harmed person, the offender, and the community." [146, p.2] Finally, a transformative justice approach goes a step further and insists that systemic injustices and violence need a bottom-up transformation of social, political, and economic conditions that perpetuate harm in the first place [41, 68].

Based on our analyses, we propose that applying these social justice frameworks can help address collective caste-based harm in online spaces. Such an approach would benefit from the following considerations: (1) We should encourage and center knowledge systems of the anti-caste movement in ways that can inform a contextually situated vocabulary of what online harm and safety entail. (2) It is critical to view technology and society as a single unit to cultivate a sense of shared accountability for how the platforms "integrate with, reshape, and sometimes harm communities" [30] — a principle that also characterizes the transformative justice approach [47]. (3) Even as we press online platforms to design and enforce nuanced community standards, it is important to recognize that these companies are ultimately driven by profit and are likely to offer only punitive measures and increased policing of social media users in response to harm. Such measures do not challenge the status quo and are often incompatible with social justice goals [46]. A careful deliberation on these issues is required as we continue to build mechanisms to address identity-based online harm.

## 6 Limitations and Future Work

Our study examining caste articulations online has certain limitations. For one, we do not know the precise caste identities of the individuals or the groups that managed these profiles. While some profiles may be governed by people from the same or adjacent caste groups, others (say, accounts fighting affirmative action) may be governed by a caste-diverse team. Since our methodological choices did not involve interactions with the groups behind the profiles, we had no insight into how caste dynamics influenced the composition of the teams and, in turn, their content. This meant that we had a window into only what was publicly available, i.e., their online caste expressions. Despite the limitations, we chose to analyze caste-positive community profiles because they bring into focus the contemporary mechanics that keep caste alive and thriving. This exercise, as such, helped us discern both blunt and subtle online caste dynamics.

Given that the authors were fluent in Hindi, Marathi, Kannada, and English, the sampled profiles were concentrated both linguistically and, to some extent, geographically. Most of the profiles we gathered were addressing Hindi-speaking audiences in northern India (states of Uttar Pradesh, Rajasthan, Madhya Pradesh, and Bihar) and parts of southern India (state of Karnataka) and western India (state of Maharashtra). Given that caste dynamics vary significantly from region to region and even from language to language, our analyses do not capture the full breadth of caste articulations in India. Yet, we believe that our qualitative analysis draws upon critical approaches that offer constructive insights into the ongoing marginalization of the lower castes by providing a thorough critique of dominant caste techniques.

The data collection and analysis were conducted between September 2023 and April 2024, when India's General Elections were imminent, the Hindu-nationalist party of BJP was set to form the government for the third consecutive time, and the Ram Mandir (a temple built on the site of the Babri Mosque, which was demolished by Hindu nationalists in 1992) was inaugurated by the prime minister in January 2024. All these socio-political events were frequently analyzed by the caste-positive accounts. While discussing the role of caste in relation to each of these events is beyond the scope of the paper, we think there is value in examining such online discourses to understand better how caste colors the current socio-political moment in India.

In addition, our study did not explore how gender and religion intersect with caste in online spaces. Though we found instances describing how women's bodies and sexualities are policed to ensure caste purity (see Section 4.2.4), a follow-up study dedicated to gendered variations in caste would be useful to shed light on how both categories intersect and contribute to the subjugation of upper and lower caste women, albeit in very different ways. In addition, even though caste is justified by drawing upon the Hindu religion, it is also practiced in other religions, including Christianity, Sikhism, and Islam in India [8, 9]. Studies focusing on caste articulations in other religions will be helpful in further understanding the malleability of caste.

Finally, this research can be complemented by studies employing other methodologies. A large-scale quantitative analysis of social media logs, such as topical analysis, could help characterize the patterns and frequencies of caste articulations online. A social network analysis could provide additional insight into how these accounts are connected and how their operations contribute to the consolidation and reproduction of caste.

## 7 Conclusion

In this paper, we aimed to deepen scholarly understanding of how caste manifests on social media platforms such as X by examining caste-positive community profiles. In particular, our goal was to surface the discursive practices that these accounts deploy in order to communicate caste status and how platform affordances support their online activities. Using a critical discourse analysis approach, we analyzed 50 caste-positive community profiles. We found that the accounts exploit platform affordances such as information control, bandwidth, visibility, searchability, and shareability to deploy both rhetorical and organizing strategies that they implement via a range of tactics. Caste-positive communities employ these tactics to construct two main arguments: (1) that their caste culture deserves a superior status and (2) that they are the "true" victims of caste discrimination. Such digitally mediated discursive practices contribute to the marginalization of lower castes by normalizing caste cultures, strengthening caste networks, and diminishing anti-caste measures. Further, this research highlights the importance of studying online activities of powerful communities, such as the dominant castes, provides theoretical frameworks and methodological suggestions to conduct such research, and argues that HCI and social computing fields should frame marginalization as a severe manifestation of online harm that demands the attention of researchers, practitioners, governments, and the civil society.

## Acknowledgments

We are grateful to the reviewers for their thoughtful engagement with our paper and for helping us improve it. We also thank Britt Paris, Krishna Akhil Kumar Adavi, Brendan Mahoney, and Jeffery Lane for their valuable feedback, and Sandy Kaplan for her assistance with editing.

## References

[1] Louis Althusser. 2001. Ideology and Ideological State Apparatuses. In *Lenin and Philosophy and Other Essays*. NYU Press, 85–126. http://www.jstor.org/stable/j.ctt9qgh9v.9

[2] Bhimrao Ramji Ambedkar. 1917. Castes In India: Their Mechanism, Genesis and Development. *Indian Antiquary* XLVI (1917). https://akscusa.org/wp-content/uploads/2020/04/lecture-2-collated-readings.pdf

[3] Derek Anderson. 2020. Linguistic Hijacking. *FPQ* 6, 3 (Sept. 2020). doi:10.5206/fpq/2020.3.8162

[4] Ira Anjali Anwar, Joyojeet Pal, and Julie Hui. 2021. Watched, but Moving: Platformization of Beauty Work and Its Gendered Mechanisms of Control. *Proc. ACM Hum.-Comput. Interact.* 4, CSCW3 (Jan. 2021), 1–20. doi:10.1145/3432949

[5] Barbara Applebaum. 2010. *Being White, Being Good: White Complicity, White Moral Responsibility, and Social Justice Pedagogy*. Lexington Books. Google-Books-ID: eJMZ0RxltAcC.

[6] Chris Atton. 2006. Far-right media on the internet: culture, discourse and power. *New Media & Society* 8, 4 (Aug. 2006), 573–587. doi:10.1177/1461444806065653

[7] Gajendran Ayyathurai. 2021. It is time for a new subfield: 'Critical Caste Studies'. https://blogs.lse.ac.uk/southasia/2021/07/05/it-is-time-for-a-new-subfield-critical-caste-studies/

[8] Shireen Azam. 2023. The political life of Muslim caste: articulations and frictions within a Pasmanda identity. *Contemporary South Asia* 31, 3 (July 2023), 426–441. doi:10.1080/09584935.2023.2237417 Publisher: Routledge _eprint: https://doi.org/10.1080/09584935.2023.2237417.

[9] Shireen Azam. 2023. Scheduled Caste Status for Dalit Muslims and Christians. *Economic and Political Weekly* 58, 27 (July 2023). https://www.epw.in/journal/2023/27/commentary/scheduled-caste-status-dalit-muslims-and.html

[10] Ramesh Bairy. 2016. *Being Brahmin, being modern: exploring the lives of caste today.* Routledge India, New Delhi.

[11] Anirban Baishya. 2015. Selfies| #NaMo: The Political Work of the Selfie in the 2014 Indian General Elections. *International Journal of Communication* 9, 0 (2015). https://ijoc.org/index.php/ijoc/article/view/3133

[12] Anirban K. Baishya. 2022. Violent spectating: Hindutva music and audio-visualizations of hate and terror in Digital India. *Communication and Critical/Cultural Studies* 19, 3 (July 2022), 289–309. doi:10.1080/14791420.2022.2099918

[13] Shelly Ghai Bajaj. 2017. The Use of Twitter during the 2014 Indian General Elections: Framing, Agenda-Setting, and the Personalization of Politics. *Asian Survey* 57, 2 (2017), 249–270. https://www.jstor.org/stable/26367749 Publisher: University of California Press.

[14] J Balasubramaniam. 2011. Dalits and a Lack of Diversity in the Newsroom. *Economic and Political Weekly* 46, 11 (2011), 21–23. https://www.jstor.org/stable/41151964 Publisher: Economic and Political Weekly.

[15] Shakuntala Banaji, Ram Bhat, Anushi Agarwal, Nihal Passanha, and Mukti Sadhana Pravin. 2019. *WhatsApp Vigilantes: An exploration of citizen reception and circulation of WhatsApp misinformation linked to mob violence in India.* Technical Report. Department of Media and Communications, LSE.

[16] Christoph Beckmann and Tom Gross. 2014. Social Computing — Bridging the Gap between the Social and the Technical. In *Social Computing and Social Media: 6th International Conference, SCSM 2014, Held as Part of HCI*

*International 2014, Heraklion, Crete, Greece, June 22-27, 2014. Proceedings*, Gabriele Meiselwitz, David Hutchison, Takeo Kanade, Josef Kittler, Jon M. Kleinberg, Alfred Kobsa, Friedemann Mattern, John C. Mitchell, Moni Naor, Oscar Nierstrasz, C. Pandu Rangan, Bernhard Steffen, Demetri Terzopoulos, Doug Tygar, and Gerhard Weikum (Eds.). Lecture Notes in Computer Science, Vol. 8531. Springer International Publishing, Cham. doi:10.1007/978-3-319-07632-4

[17] Rosanna Bellini, Debora De Castro Leal, Hazel Anneke Dixon, Sarah E Fox, and Angelika Strohmayer. 2022. "There is no justice, just us": Making mosaics of justice in social justice Human-Computer Interaction. In *CHI Conference on Human Factors in Computing Systems Extended Abstracts*. ACM, New Orleans LA USA, 1–6. doi:10.1145/3491101.3503698

[18] Gabrielle Benabdallah, Michael W. Beach, Nathanael Elias Mengist, Daniela Rosner, Kavita S Philip, and Lucy Suchman. 2023. The Politics of Imaginaries: Probing Humanistic Inquiry in HCI. In *Designing Interactive Systems Conference*. ACM, Pittsburgh PA USA, 131–134. doi:10.1145/3563703.3591457

[19] Kiran Vinod Bhatia. 2022. Hindu Nationalism Online: Twitter as Discourse and Interface. *Religions* 13, 8 (Aug. 2022), 739. doi:10.3390/rel13080739

[20] Ayesha Bhimdiwala, Krishna Akhil Kumar Adavi, and Ahmer Arif. 2024. Fighting for Their Voice: Understanding Indian Muslim Women's Responses to Networked Harassment. *Proc. ACM Hum.-Comput. Interact.* 8, CSCW1 (April 2024), 166:1–166:24. doi:10.1145/3641005

[21] Malavika Binny. 2022. Tracing the Contours of Hate Speech in India in the Pandemic Year: The Curious Case of Online Hate Speech against Muslims and Dalits During the Pandemic. *Contemporary Voice of Dalit* (May 2022), 2455328X221094364. doi:10.1177/2455328X221094364 Publisher: SAGE Publications India.

[22] Pierre Bourdieu. 2005. Habitus. In *Habitus: a sense of place* (2nd ed ed.), Jean Hillier and Emma Rooksby (Eds.). Ashgate, Aldershot, Hants, England ; Burlington, VT.

[23] John W. Bowers, Donovan J. Ochs, Richard J. Jensen, and David P. Schulz. 2009. *The Rhetoric of Agitation and Control: Third Edition*. Waveland Press. Google-Books-ID: ta4QAAAAQBAJ.

[24] danah boyd and J. Heer. 2006. Profiles as Conversation: Networked Identity Performance on Friendster. In *Proceedings of the 39th Annual Hawaii International Conference on System Sciences (HICSS'06)*. IEEE, Kauia, HI, USA, 59c–59c. doi:10.1109/HICSS.2006.394

[25] Virginia Braun and Victoria Clarke. 2022. *Thematic Analysis*. SAGE Publications.

[26] Nicola Bruno, Giorgia Guerra, Brigitta Pia Alioto, and Alessandra Cecilia Jacomuzzi. 2023. Shareability: novel perspective on human-media interaction. *Front. Comput. Sci.* 5 (Sept. 2023). doi:10.3389/fcomp.2023.1106322 Publisher: Frontiers.

[27] André Béteille. 2012. The Peculiar Tenacity of Caste. *Economic and Political Weekly* 47, 13 (2012), 41–48. https://www.jstor.org/stable/23214709 Publisher: Economic and Political Weekly.

[28] Justine Calma. 2023. Scientists say they can't rely on Twitter anymore. https://www.theverge.com/2023/5/31/23739084/twitter-elon-musk-api-policy-chilling-academic-research

[29] Eshwar Chandrasekharan, Shagun Jhaver, Amy Bruckman, and Eric Gilbert. 2022. Quarantined! Examining the Effects of a Community-Wide Moderation Intervention on Reddit. *ACM Trans. Comput.-Hum. Interact.* 29, 4, Article 29 (March 2022), 26 pages. doi:10.1145/3490499

[30] Brian J Chen and Jacob Metcalf. 2024. Explainer: A Sociotechnical Approach to AI Policy. (2024).

[31] Marika Cifor and Patricia Garcia. 2019. Gendered by Design: A Duoethnographic Study of Personal Fitness Tracking Systems. *Trans. Soc. Comput.* 2, 4 (Dec. 2019), 1–22. doi:10.1145/3364685

[32] Scott Counts, Munmun De Choudhury, Jana Diesner, Eric Gilbert, Marta Gonzalez, Brian Keegan, Mor Naaman, and Hanna Wallach. 2014. Computational social science: CSCW in the social media era. In *Proceedings of the companion publication of the 17th ACM conference on Computer supported cooperative work & social computing*. ACM, Baltimore Maryland USA, 105–108. doi:10.1145/2556420.2556849

[33] Blyth Crawford, Florence Keen, and Guillermo Suarez de Tangil. 2020. *Memetic Irony and the Promotion of Violence within Chan Cultures*. Technical Report. Centre for Research and Evidence on Security Threats.

[34] Valentine Crosset, Samuel Tanner, and Aurélie Campana. 2019. Researching far right groups on Twitter: Methodological challenges 2.0. *New Media & Society* 21, 4 (April 2019), 939–961. doi:10.1177/1461444818817306

[35] Brittany I. Davidson, Darja Wischerath, Daniel Racek, Douglas A. Parry, Emily Godwin, Joanne Hinds, Dirk Van Der Linden, Jonathan F. Roscoe, Laura Ayravainen, and Alicia G. Cork. 2023. Platform-controlled social media APIs threaten open science. *Nat Hum Behav* 7, 12 (Nov. 2023), 2054–2057. doi:10.1038/s41562-023-01750-2

[36] Munmun De Choudhury, Shagun Jhaver, Benjamin Sugar, and Ingmar Weber. 2016. Social media participation in an activist movement for racial equality. In *Proceedings of the international aaai conference on web and social media*, Vol. 10. 92–101.

[37] Satish Deshpande. 2006. Exclusive Inequalities: Merit, Caste and Discrimination in Indian Higher Education Today. *Economic and Political Weekly* 41, 24 (2006), 2438–2444. https://www.jstor.org/stable/4418346 Publisher: Economic

and Political Weekly.

[38] Ahmad Diab, Bolor-Erdene Jagdagdorj, Lynnette Hui Xian Ng, Yu-Ru Lin, and Michael Miller Yoder. 2023. Online to Offline Crossover of White Supremacist Propaganda. In *Companion Proceedings of the ACM Web Conference 2023 (WWW '23 Companion)*. Association for Computing Machinery, New York, NY, USA, 1308–1316. doi:10.1145/3543873.3587569

[39] Catherine D'Ignazio and Lauren F. Klein. 2020. *Data feminism.* The MIT Press, Cambridge, Massachusetts.

[40] Claire Stravato Emes and Arul Chib. 2022. Co-opted Marginality in a Controlled Media Environment: The Influence of Social Media Affordances on the Immigration Discourse. *Trans. Soc. Comput.* 5, 1-4 (Dec. 2022), 1–15. doi:10.1145/3532103

[41] Sheena Erete, Karla Thomas, Denise Nacu, Jessa Dickinson, Naomi Thompson, and Nichole Pinkard. 2021. Applying a Transformative Justice Approach to Encourage the Participation of Black and Latina Girls in Computing. *ACM Trans. Comput. Educ.* 21, 4 (Dec. 2021), 1–24. doi:10.1145/3451345

[42] Sandra K. Evans, Katy E. Pearce, Jessica Vitak, and Jeffrey W. Treem. 2017. Explicating Affordances: A Conceptual Framework for Understanding Affordances in Communication Research. *J Comput-Mediat Comm* 22, 1 (Jan. 2017), 35–52. doi:10.1111/jcc4.12180

[43] Norman Fairclough. 1995. *Critical discourse analysis: the critical study of language.* Longman, Harlow, England.

[44] Tracie Farrell, Oscar Araque, Miriam Fernandez, and Harith Alani. 2020. On the use of Jargon and Word Embeddings to Explore Subculture within the Reddit's Manosphere. In *12th ACM Conference on Web Science*. ACM, Southampton United Kingdom, 221–230. doi:10.1145/3394231.3397912

[45] Tracie Farrell, Miriam Fernandez, Jakub Novotny, and Harith Alani. 2019. Exploring Misogyny across the Manosphere in Reddit. In *Proceedings of the 10th ACM Conference on Web Science*. ACM, Boston Massachusetts USA, 87–96. doi:10.1145/3292522.3326045

[46] Andrew Feenberg. 2010. *Transforming technology: a critical theory revisited* (im kolophon: 2010 ed.). Oxford Univ. Press, New York, NY.

[47] Barnard Center for Research on Women. 2020. What is Transformative Justice? https://bcrw.barnard.edu/videos/what-is-transformative-justice/

[48] Jesse Fox and Bree McEwan. 2017. Distinguishing technologies for social interaction: The perceived social affordances of communication channels scale. *Communication Monographs* 84, 3 (July 2017), 298–318. doi:10.1080/03637751.2017.1332418

[49] Sarah Fox, Mariam Asad, Katherine Lo, Jill P. Dimond, Lynn S. Dombrowski, and Shaowen Bardzell. 2016. Exploring Social Justice, Design, and HCI. In *Proceedings of the 2016 CHI Conference Extended Abstracts on Human Factors in Computing Systems (CHI EA '16)*. Association for Computing Machinery, New York, NY, USA, 3293–3300. doi:10.1145/2851581.2856465

[50] Deen Freelon. 2018. Computational research in the post-API age. *Political Communication* 35, 4 (2018), 665–668.

[51] Carolyn Gallaher. 2021. Mainstreaming white supremacy: a twitter analysis of the American 'Alt-Right'. *Gender, Place & Culture* 28, 2 (Feb. 2021), 224–252. doi:10.1080/0966369X.2019.1710472 Publisher: Routledge _eprint: https://doi.org/10.1080/0966369X.2019.1710472.

[52] Jasmine K Gani and Rabea M Khan. 2024. Positionality Statements as a Function of Coloniality: Interrogating Reflexive Methodologies. *International Studies Quarterly* 68, 2 (June 2024), sqae038. doi:10.1093/isq/sqae038

[53] Piyush Ghasiya, Georg Ahnert, and Kazutoshi Sasahara. 2023. Identifying Themes of Right-Wing Extremism in Hindutva Discourse on Twitter. *Social Media* (2023).

[54] James J. Gibson. 1979. The Theory of Affordances. In *The ecological approach to visual perception.* Houghton Mifflin, Boston, 118–135.

[55] Debbie Ging. 2019. Alphas, Betas, and Incels: Theorizing the Masculinities of the Manosphere. *Men and Masculinities* 22, 4 (Oct. 2019), 638–657. doi:10.1177/1097184X17706401 Publisher: SAGE Publications Inc.

[56] Lisa Given. 2012. *Looking for Information.* Vol. 3. Emerald Group Publishing, Bingley. doi:10.1108/S1876-0562(2012)002012b003

[57] Erving Goffman. 1959. *The Presentation of Self in Everyday Life.* Knopf Doubleday Publishing Group. https://books.google.co.in/books?id=Sdt-cDkV8pQC

[58] Nitin Govil and Anirban Kapil Baishya. 2018. The Bully in the Pulpit: Autocracy, Digital Social Media, and Right-wing Populist Technoculture. *Communication, Culture and Critique* 11, 1 (March 2018), 67–84. doi:10.1093/ccc/tcx001

[59] Gopal Guru. 2012. Experience, Space, and Justice. In *The Cracked Mirror: An Indian Debate on Experience and Theory.* OUP India. https://books.google.co.in/books?id=BFfOygAACAAJ

[60] Gopal Guru and Sundar Sarukkai. 2012. *The Cracked Mirror: An Indian Debate on Experience and Theory.* OUP India. https://books.google.co.in/books?id=BFfOygAACAAJ

[61] Eunbin Ha, Haein Kong, and Shagun Jhaver. 2025. Examining racial stereotypes in YouTube autocomplete suggestions. *New Media & Society* (2025), 14614448251346503.

[62] Hussam Habib, Padmini Srinivasan, and Rishab Nithyanand. 2022. Making a Radical Misogynist: How Online Social Engagement with the Manosphere Influences Traits of Radicalization. *Proc. ACM Hum.-Comput. Interact.* 6, CSCW2 (Nov. 2022), 1–28. doi:10.1145/3555551

[63] Loni Hagen, Mary Falling, Oleksandr Lisnichenko, AbdelRahim A. Elmadany, Pankti Mehta, Muhammad Abdul-Mageed, Justin Costakis, and Thomas E. Keller. 2019. Emoji Use in Twitter White Nationalism Communication. In *Companion Publication of the 2019 Conference on Computer Supported Cooperative Work and Social Computing (CSCW '19 Companion)*. Association for Computing Machinery, New York, NY, USA, 201–205. doi:10.1145/3311957.3359495

[64] Stuart Hall, Jessica Evans, and Sean Nixon (Eds.). 2013. *Representation: Cultural Representations and Signifying Practices* (second edition ed.). Sage ; The Open University, Los Angeles : Milton Keynes, United Kingdom. OCLC: ocn842411043.

[65] Tejas Harad. 2018. Towards an internet of equals. https://lifestyle.livemint.com/news/talking-point/towards-an-internet-of-equals-111645092715631.html

[66] Bing He, Caleb Ziems, Sandeep Soni, Naren Ramakrishnan, Diyi Yang, and Srijan Kumar. 2021. Racism is a virus: anti-asian hate and counterspeech in social media during the COVID-19 crisis. In *Proceedings of the 2021 IEEE/ACM International Conference on Advances in Social Networks Analysis and Mining*. ACM, Virtual Event Netherlands, 90–94. doi:10.1145/3487351.3488324

[67] Øyvind Ihlen and Robert L. Heath. 2018. Conclusions and Take Away Points. In *The Handbook of Organizational Rhetoric and Communication* (1 ed.), Øyvind Ihlen and Robert L. Heath (Eds.). Wiley. doi:10.1002/9781119265771

[68] Walidah Imarisha, Alexis Gumbs, Leah Lakshmi Piepzna-Samarasinha, Adrienne Maree Brown, and Mia Mingus. 2017. The Fictions and Futures of Transformative Justice. https://thenewinquiry.com/the-fictions-and-futures-of-transformative-justice/

[69] ICF International Institute for Population Sciences (IIPS). 2021. *National Family Health Survey -5 (2019-21)*. Technical Report Volume 1. Mumbai. http://rchiips.org/nfhs/NFHS-5Reports/NFHS-5_INDIA_REPORT.pdf

[70] Heesoo Jang, Nanditha Narayanamoorthy, Laura Schelenz, Lou Therese Brandner, Anne Burkhardt, Simon David Hirsbrunner, Jessica Pidoux, Scott Timcke, Airi Lampinen, and Riyaj Isamiya Shaikh. 2023. Platform (In)Justice: Exploring Research Priorities and Practical Solutions. In *Computer Supported Cooperative Work and Social Computing*. ACM, Minneapolis MN USA, 576–580. doi:10.1145/3584931.3606953

[71] Henry Jenkins. 2006. *Fans, Bloggers, and Gamers: Exploring Participatory Culture.* NYU Press. https://books.google.sm/books?id=-gcLB-7FkBQC

[72] Shagun Jhaver. 2025. Bans vs. Warning Labels: Examining Bystanders' Support for Community-wide Moderation Interventions. *ACM Trans. Comput.-Hum. Interact.* 32, 2, Article 18 (April 2025), 30 pages. doi:10.1145/3715116

[73] Shagun Jhaver, Darren Scott Appling, Eric Gilbert, and Amy Bruckman. 2019. "Did You Suspect the Post Would Be Removed?": Understanding User Reactions to Content Removals on Reddit. *Proc. ACM Hum.-Comput. Interact.* 3, CSCW, Article 192 (Nov. 2019), 33 pages. doi:10.1145/3359294

[74] Shagun Jhaver, Iris Birman, Eric Gilbert, and Amy Bruckman. 2019. Human-Machine Collaboration for Content Regulation: The Case of Reddit Automoderator. *ACM Trans. Comput.-Hum. Interact.* 26, 5, Article 31 (July 2019), 35 pages. doi:10.1145/3338243

[75] Shagun Jhaver, Christian Boylston, Diyi Yang, and Amy Bruckman. 2021. Evaluating the Effectiveness of Deplatforming as a Moderation Strategy on Twitter. *Proc. ACM Hum.-Comput. Interact.* 5, CSCW2, Article 381 (oct 2021), 30 pages. doi:10.1145/3479525

[76] Shagun Jhaver, Amy Bruckman, and Eric Gilbert. 2019. Does Transparency in Moderation Really Matter? User Behavior After Content Removal Explanations on Reddit. *Proc. ACM Hum.-Comput. Interact.* 3, CSCW, Article 150 (nov 2019), 27 pages. doi:10.1145/3359252

[77] Shagun Jhaver, Larry Chan, and Amy Bruckman. 2018. The View from the Other Side: The Border Between Controversial Speech and Harassment on Kotaku in Action. *First Monday* 23, 2 (2018). http://firstmonday.org/ojs/index.php/fm/article/view/8232

[78] Shagun Jhaver, Quan Ze Chen, Detlef Knauss, and Amy X. Zhang. 2022. Designing Word Filter Tools for Creator-Led Comment Moderation. In *Proceedings of the 2022 CHI Conference on Human Factors in Computing Systems* (New Orleans, LA, USA) *(CHI '22)*. Association for Computing Machinery, New York, NY, USA, Article 205, 21 pages. doi:10.1145/3491102.3517505

[79] Shagun Jhaver, Sucheta Ghoshal, Amy Bruckman, and Eric Gilbert. 2018. Online Harassment and Content Moderation: The Case of Blocklists. *ACM Trans. Comput.-Hum. Interact.* 25, 2, Article 12 (March 2018), 33 pages. doi:10.1145/3185593

[80] Shagun Jhaver, Himanshu Rathi, and Koustuv Saha. 2024. Bystanders of online moderation: Examining the effects of witnessing post-removal explanations. In *Proceedings of the 2024 CHI Conference on Human Factors in Computing Systems*. 1–9.

[81] Shagun Jhaver, Alice Qian Zhang, Quan Ze Chen, Nikhila Natarajan, Ruotong Wang, and Amy X. Zhang. 2023. Personalizing Content Moderation on Social Media: User Perspectives on Moderation Choices, Interface Design, and Labor. 7, CSCW2, Article 289 (Oct. 2023), 33 pages. doi:10.1145/3610080

[82] Damni Kain, Shivangi Narayan, Torsha Sarkar, and Gurshabad Grover. 2021. Online caste-hate speech: Pervasive discrimination and humiliation on social media.

[83] Azade E. Kakavand. 2024. Far-right social media communication in the light of technology affordances: a systematic literature review. *Annals of the International Communication Association* 48, 1 (Jan. 2024), 37–56. doi:10.1080/23808985.2023.2280824

[84] Victor Kaptelinin and Bonnie Nardi. 2012. Affordances in HCI: toward a mediated action perspective. In *Proceedings of the SIGCHI Conference on Human Factors in Computing Systems*. ACM, Austin Texas USA, 967–976. doi:10.1145/2207676.2208541

[85] Nayana Kirasur. 2021. How Has the Pandemic Deepened Inequalities in India's School Education? https://itforchange.net/has-pandemic-deepened-inequalities-india%E2%80%99s-school-education

[86] Nayana Kirasur. 2021. What Our Textbooks Aren't Telling Us About Caste in India. *THE BASTION* (June 2021). https://thebastion.co.in/politics-and/education/syllabus-and-pedagogy/what-our-textbooks-arent-telling-us-about-caste-in-india/

[87] Nayana Kirasur. 2022. How Does Aadhaar-Enabled Welfare Delivery Exclude Women? https://itforchange.net/notes-from-field-how-does-aadhaar-enabled-welfare-delivery-exclude-women

[88] Nayana Kirasur. 2022. Recording the Unorganized Sector: Reflections on the e-Shram Database. https://itforchange.net/notes-from-field-recording-unorganized-sector-reflections-on-e-shram-database

[89] Nayana Kirasur. 2022. What it Takes to Maintain Access to the Digital: Experiences of Adolescent Girls in Rural Mysuru. https://itforchange.net/notes-from-field-what-it-takes-to-maintain-access-to-digital-experiences-of-adolescent-girls-rural

[90] Nayana Kirasur. 2022. Wielding Informational Power as a Feminist Practice: Vignettes from Rural Mysuru. https://itforchange.net/notes-from-field-wielding-informational-power-as-a-feminist-practice-vignettes-from-rural-mysuru

[91] Ramaravind Kommiya Mothilal, Dibyendu Mishra, Sachita Nishal, Faisal M. Lalani, and Joyojeet Pal. 2022. Voting with the Stars: Analyzing Partisan Engagement between Celebrities and Politicians in India. *Proc. ACM Hum.-Comput. Interact.* 6, CSCW1 (March 2022), 1–29. doi:10.1145/3512981

[92] Vivek Kumar. 2016. Caste, Contemporaneity and Assertion. *Economic and Political Weekly* 51, 50 (2016), 84–86. http://www.jstor.org.proxy.libraries.rutgers.edu/stable/44165968 Publisher: Economic and Political Weekly.

[93] Michèle Lamont. 2017. Prisms of Inequality: Moral Boundaries, Exclusion, and Academic Evaluation. In *Praemium Erasmianum Essay 2017*. Praemium Erasmianum Foundation, Amsterdam. Backup Publisher: Praemium Erasmianum Foundation.

[94] Christopher J Lebron. 2023. *The making of black lives matter: A brief history of an idea.* Oxford University Press.

[95] Henri Lefebvre. 1997. *The production of space* (reprinted ed.). Blackwell, Oxford.

[96] Rebecca Lewis. 2018. Alternative Influence: Broadcasting the Reactionary Right on YouTube. *Data & Society* (2018).

[97] Purnima Mankekar and Hannah Carlan. 2019. The Remediation of Nationalism: Viscerality, Virality, and Digital Affect. In *Global Digital Cultures: Perspectives from South Asia*, Aswin Punathambekar and Sriram Mohan (Eds.). University of Michigan Press. doi:10.3998/mpub.9561751

[98] Rachel Marks and Mel Stanfill. 2023. Methodological Solutions for the Challenges of Studying Racist Communication on Social Media. In *Proceedings of the 41st ACM International Conference on Design of Communication*. ACM, Orlando FL USA, 70–76. doi:10.1145/3615335.3623013

[99] Alice E. Marwick. 2021. Morally Motivated Networked Harassment as Normative Reinforcement. *Social Media + Society* 7, 2 (April 2021), 20563051211021378. doi:10.1177/20563051211021378 Publisher: SAGE Publications Ltd.

[100] Smeeta Mishra and Krishna Jayakar. 2019. Remarriage in India: Online Presentation Strategies of Men and Women on an Indian Remarriage Website.

[101] Aparna Moitra, Syed Ishtiaque Ahmed, and Priyank Chandra. 2021. Parsing the 'Me' in #MeToo: Sexual Harassment, Social Media, and Justice Infrastructures. *Proc. ACM Hum.-Comput. Interact.* 5, CSCW1 (April 2021), 111:1–111:34. doi:10.1145/3449185

[102] Rahul Mukherjee and Fathima Nizaruddin. 2022. Digital Platforms in Contemporary India: The Transformation of Quotidian Life Worlds. *Asiascape: Digital Asia* 9, 1-2 (July 2022), 5–18. doi:10.1163/22142312-bja10026 Publisher: Brill.

[103] Nandan Nilekani. 2009. *Imagining India: the Idea of a Renewed Nation.* Penguin Books, New York. OCLC: 883350221.

[104] Gail Omvedt. 1993. *Reinventing revolution: new social movements and the socialist tradition in India.* M.E. Sharpe, Armonk, N.Y.

[105] Gail Omvedt. 2004. Untouchables in the world of IT. http://panoslondon.panosnetwork.org/features/untouchables-in-the-world-of-it/

[106] Zizi Papacharissi. 2009. The virtual sphere 2.0 The internet, the public sphere, and beyond. In *Routledge Handbook of Internet Politics*, Andrew Chadwick and Philip N. Howard (Eds.). Routledge, London ; New York. OCLC: ocn191318084.

[107] Kostantinos Papadamou, Savvas Zannettou, Jeremy Blackburn, Emiliano De Cristofaro, Gianluca Stringhini, and Michael Sirivianos. 2021. "How over is it?" Understanding the Incel Community on YouTube. *Proc. ACM Hum.-Comput. Interact.* 5, CSCW2 (Oct. 2021), 412:1–412:25. doi:10.1145/3479556

[108] Gaurav J. Pathania and William G. Tierney. 2018. An ethnography of caste and class at an Indian university: creating capital. *Tertiary Education and Management* (Feb. 2018), 1–11. doi:10.1080/13583883.2018.1439998

[109] Ashwin Rajadesingan, Ramaswami Mahalingam, and David Jurgens. 2019. Smart, Responsible, and Upper Caste Only: Measuring Caste Attitudes through Large-Scale Analysis of Matrimonial Profiles. *ICWSM* 13 (July 2019), 393–404. doi:10.1609/icwsm.v13i01.3239

[110] Saranya Rajiakodi, Bharathi Raja Chakravarthi, Rahul Ponnusamy, Prasanna Kumaresan, Sathiyaraj Thangasamy, Bhuvaneswari Sivagnanam, and Charmathi Rajkumar. 2024. Overview of Shared Task on Caste and Migration Hate Speech Detection. In *Proceedings of the Fourth Workshop on Language Technology for Equality, Diversity, Inclusion*, Bharathi Raja Chakravarthi, Bharathi B, Paul Buitelaar, Thenmozhi Durairaj, György Kovács, and Miguel Ángel García Cumbreras (Eds.). Association for Computational Linguistics, St. Julian's, Malta, 145–151. https://aclanthology.org/2024.ltedi-1.14

[111] Shakuntala Rao. 2018. Making of Selfie Nationalism: Narendra Modi, the Paradigm Shift to Social Media Governance, and Crisis of Democracy. *Journal of Communication Inquiry* 42, 2 (April 2018), 166–183. doi:10.1177/0196859917754053 Publisher: SAGE Publications Inc.

[112] Devanshu Sajlan. 2021. Hate Speech against Dalits on Social Media: Would a Penny Sparrow be Prosecuted in India for Online Hate Speech? *CASTE: A Global Journal on Social Exclusion* 2, 1 (2021), 77–96. https://www.jstor.org/stable/48643386 Publisher: Brandeis University, Center for Global Development and Sustainability.

[113] Nithya Sambasivan, Amna Batool, Nova Ahmed, Tara Matthews, Kurt Thomas, Laura Sanely Gaytán-Lugo, David Nemer, Elie Bursztein, Elizabeth Churchill, and Sunny Consolvo. 2019. "They Don't Leave Us Alone Anywhere We Go": Gender and Digital Abuse in South Asia. In *Proceedings of the 2019 CHI Conference on Human Factors in Computing Systems (CHI '19)*. Association for Computing Machinery, New York, NY, USA, 1–14. doi:10.1145/3290605.3300232

[114] Sebastian Schelter and Jérôme Kunegis. 2017. 'Dark Germany': Temporal Characteristics and Connectivity Patterns in Online Far-Right Protests Against Refugee Housing. In *Proceedings of the 2017 ACM on Web Science Conference (WebSci '17)*. Association for Computing Machinery, New York, NY, USA, 415–416. doi:10.1145/3091478.3098880

[115] Morgan Klaus Scheuerman, Stacy M. Branham, and Foad Hamidi. 2018. Safe Spaces and Safe Places: Unpacking Technology-Mediated Experiences of Safety and Harm with Transgender People. *Proc. ACM Hum.-Comput. Interact.* 2, CSCW (Nov. 2018), 1–27. doi:10.1145/3274424

[116] Morgan Klaus Scheuerman, Jialun Aaron Jiang, Casey Fiesler, and Jed R. Brubaker. 2021. A Framework of Severity for Harmful Content Online. *Proc. ACM Hum.-Comput. Interact.* 5, CSCW2 (Oct. 2021), 1–33. doi:10.1145/3479512 arXiv:2108.04401 [cs].

[117] Phoebe Sengers, John McCarthy, and Paul Dourish. 2006. Reflective HCI: articulating an agenda for critical practice. In *CHI '06 Extended Abstracts on Human Factors in Computing Systems (CHI EA '06)*. Association for Computing Machinery, New York, NY, USA, 1683–1686. doi:10.1145/1125451.1125762

[118] Juan Carlos Medina Serrano, Morteza Shahrezaye, Orestis Papakyriakopoulos, and Simon Hegelich. 2019. The Rise of Germany's AfD: A Social Media Analysis. In *Proceedings of the 10th International Conference on Social Media and Society*. ACM, Toronto ON Canada, 214–223. doi:10.1145/3328529.3328562

[119] Ghanshyam Shah. 1991. Social Backwardness and Politics of Reservations. *Economic and Political Weekly* 26, 11/12 (1991), 601–610. https://www.jstor.org/stable/4397417 Publisher: Economic and Political Weekly.

[120] Farhana Shahid and Aditya Vashistha. 2023. Decolonizing Content Moderation: Does Uniform Global Community Standard Resemble Utopian Equality or Western Power Hegemony?. In *Proceedings of the 2023 CHI Conference on Human Factors in Computing Systems (CHI '23)*. Association for Computing Machinery, New York, NY, USA, 1–18. doi:10.1145/3544548.3581538

[121] Murali Shanmugavelan. 2022. The Case for Critical Caste and Technology Studies. https://medium.com/datasociety-points/the-case-for-critical-caste-and-technology-studies-b987dcf20c8d

[122] Murali Shanmugavelan. 2022. Caste-hate speech and digital media politics. *Journal of Digital Media & Policy* 13, 1 (March 2022), 41–55. doi:10.1386/jdmp_00089_1

[123] Clay Shirky. 2011. The Political Power of Social Media: Technology, the Public Sphere, and Political Change. *Foreign Affairs* 90, 1 (2011), 28–41. https://www.jstor.org/stable/25800379 Publisher: Council on Foreign Relations.

[124] Divyanshu Kumar Singh and Palashi Vaghela. 2024. Anti-Caste Lessons for Computing: Educate, Agitate, Organize. *XRDS* 30, 4 (June 2024), 41–45. doi:10.1145/3665600

[125] Santosh K Singh. 2017. The Caste Question and Songs of Protest in Punjab. *Economic and Political Weekly* 52, 34 (2017), 33–37. https://www.jstor.org/stable/26695745 Publisher: Economic and Political Weekly.

[126] T. Soundararajan. 2019. *Facebook India: towards the tipping point of violence: caste and religious hate speech.* Equality Labs, USA.

[127] Megan Squire. 2021. Monetizing Propaganda: How Far-right Extremists Earn Money by Video Streaming. In *Proceedings of the 13th ACM Web Science Conference 2021 (WebSci '21)*. Association for Computing Machinery, New York, NY, USA, 158–167. doi:10.1145/3447535.3462490

[128] M. N. Srinivas. 1956. A Note on Sanskritization and Westernization. *The Far Eastern Quarterly* 15, 4 (Aug. 1956), 481. doi:10.2307/2941919

[129] James Stevenson, Matthew Edwards, and Awais Rashid. 2024. Analysing The Activities Of Far-Right Extremists On The Parler Social Network. In *Proceedings of the 2023 IEEE/ACM International Conference on Advances in Social Networks Analysis and Mining (ASONAM '23)*. Association for Computing Machinery, New York, NY, USA, 392–399. doi:10.1145/3625007.3627733

[130] Arvind Kumar Thakur. 2020. New Media and the Dalit Counter-public Sphere. *Television & New Media* 21, 4 (May 2020), 360–375. doi:10.1177/1527476419872133 Publisher: SAGE Publications.

[131] Michaelanne Thomas, Neha Kumar, Ari Schlesinger, Marisol Wong-Villacres, Morgan G. Ames, Rajesh Veeraraghavan, Jacki O'Neill, Joyojeet Pal, and Mary L. Gray. 2018. Solidarity Across Borders: Navigating Intersections Towards Equity and Inclusion. In *Companion of the 2018 ACM Conference on Computer Supported Cooperative Work and Social Computing*. ACM, Jersey City NJ USA, 487–494. doi:10.1145/3272973.3273007

[132] Sukhadeo Thorat and Katherine S. Newman. 2007. Caste and Economic Discrimination: Causes, Consequences and Remedies. *Economic and Political Weekly* 42, 41 (2007), 4121–4124. https://www.jstor.org/stable/40276545 Publisher: Economic and Political Weekly.

[133] Charles Tilly. 1986. *The Contentious French: 4 centuries of popular struggle.* The Belknap Pr. of Harvard Univ. Pr, Cambridge, Mass. u.a.

[134] Sahana Udupa. 2019. India Needs a Fresh Strategy to Tackle Online Extreme Speech. *Economic and Political Weekly (Engage)* 54, 4 (2019).

[135] Palashi Vaghela, Steven J Jackson, and Phoebe Sengers. 2022. Interrupting Merit, Subverting Legibility: Navigating Caste In 'Casteless' Worlds of Computing. In *CHI Conference on Human Factors in Computing Systems*. ACM, New Orleans LA USA, 1–20. doi:10.1145/3491102.3502059

[136] Palashi Vaghela, Ramaravind K Mothilal, and Joyojeet Pal. 2021. Birds of a Caste - How Caste Hierarchies Manifest in Retweet Behavior of Indian Politicians. *Proc. ACM Hum.-Comput. Interact.* 4, CSCW3 (Jan. 2021), 1–24. doi:10.1145/3432911

[137] Palashi Vaghela, Ramaravind Kommiya Mothilal, Daniel Romero, and Joyojeet Pal. 2022. Caste Capital on Twitter: A Formal Network Analysis of Caste Relations among Indian Politicians. *Proc. ACM Hum.-Comput. Interact.* 6, CSCW1 (March 2022), 1–29. doi:10.1145/3512927

[138] Divya Vaid. 2014. Caste in Contemporary India: Flexibility and Persistence. *Annu. Rev. Sociol.* 40, 1 (July 2014), 391–410. doi:10.1146/annurev-soc-071913-043303

[139] Teun A. Van Dijk. 2012. *Ideology and Discourse.* Pompeu Fabra University. https://discourses.org/wp-content/uploads/2022/06/Teun-A.-van-Dijk-2012-Ideology-And-Discourse.pdf

[140] Teun A. Van Dijk. 2015. Critical Discourse Analysis. In *The handbook of discourse analysis* (2nd edition ed.), Deborah Tannen, Heidi Ehernberger Hamilton, and Deborah Schiffrin (Eds.). Wiley Blackwell, Malden, MA Chichester, West Sussex.

[141] Alyvia Walters, Tawfiq Ammari, Kiran Garimella, and Shagun Jhaver. 2024. Online knowledge production in polarized political memes: The case of critical race theory. *New Media & Society* (2024), 14614448241252591.

[142] Ruth Wodak. 2015. Critical Discourse Analysis, Discourse-Historical Approach. In *The International Encyclopedia of Language and Social Interaction* (1 ed.), Karen Tracy, Todd Sandel, and Cornelia Ilie (Eds.). Wiley, 1–14. doi:10.1002/9781118611463.wbielsi116

[143] Ruth Wodak and Michael Meyer. 2001. *Methods of Critical Discourse Analysis.* SAGE Publications, Ltd, 6 Bonhill Street, London EC2A 4PU. doi:10.4135/9780857028020

[144] Fan Wu, Sanyam Lakhanpal, Qian Li, Kookjin Lee, Doowon Kim, Heewon Chae, and Kyounghee Hazel Kwon. 2024. Not All Asians are the Same: A Disaggregated Approach to Identifying Anti-Asian Racism in Social Media. In *Proceedings of the ACM Web Conference 2024 (WWW '24)*. Association for Computing Machinery, New York, NY, USA, 2615–2626. doi:10.1145/3589334.3645630

[145] Sijia Xiao, Coye Cheshire, and Niloufar Salehi. 2022. Sensemaking, Support, Safety, Retribution, Transformation: A Restorative Justice Approach to Understanding Adolescents' Needs for Addressing Online Harm. In *Proceedings of*

*the 2022 CHI Conference on Human Factors in Computing Systems (CHI '22)*. Association for Computing Machinery, New York, NY, USA, 1–15. doi:10.1145/3491102.3517614

[146] Sijia Xiao, Shagun Jhaver, and Niloufar Salehi. 2023. Addressing Interpersonal Harm in Online Gaming Communities: The Opportunities and Challenges for a Restorative Justice Approach. *ACM Trans. Comput.-Hum. Interact.* 30, 6 (Sept. 2023), 83:1–83:36. doi:10.1145/3603625

[147] Mary B. Ziskin. 2019. Critical discourse analysis and critical qualitative inquiry: data analysis strategies for enhanced understanding of inference and meaning. *International Journal of Qualitative Studies in Education* 32, 6 (July 2019), 606–631. doi:10.1080/09518398.2019.1609118 Publisher: Routledge _eprint: https://doi.org/10.1080/09518398.2019.1609118.